\documentclass[journal]{IEEEtran}
\usepackage[cmex10]{amsmath}
\usepackage{amssymb}
\usepackage{algorithm}
\usepackage{algorithmic}
\usepackage{graphicx}
\usepackage{cite}
\usepackage{color}
\usepackage{multirow}
\usepackage{booktabs}
\DeclareMathOperator*{\argmin}{\arg\!\min}
\usepackage{lscape}

\ifCLASSINFOpdf
\else
\fi
\newtheorem{defi}{Definition}
\newtheorem{theorem}{Theorem}
\newtheorem{lemma}{Lemma}
\newtheorem{remark}{Remark}
\newtheorem{corollary}{Corollary}

\begin{document}
\title{Power Allocation in the Energy Harvesting Full-Duplex Gaussian Relay Channels}

\author{Mahmood~Mohassel~Feghhi,
        Mahtab~Mirmohseni,
        and~Aliazam~Abbasfar,~\IEEEmembership{Senior~Member,~IEEE}% <-this % stops a space
\thanks{M. Mohassel Feghhi and A. Abbasfar are with the School of Electrical and Computer Eng., College of Eng., University of Tehran, Tehran 14395-515, IRAN (e-mail: {mohasselfeghhi, abbasfar}@ut.ac.ir).}% <-this % stops a space
\thanks{M. Mirmohseni is with the Department of Electrical Engineering, Sharif University, Tehran, IRAN (e-mail: mirmohseni@sharif.edu).}}
\maketitle

\begin{abstract}
%\boldmath
In this paper, we propose a general model to study the full-duplex non-coherent decode-and-forward Gaussian relay channel with energy harvesting (EH) nodes, called NC-EH-$\mathcal{RC}$, in three cases: $i)$ no energy transfer (ET), $ii)$ one-way ET from the source (S) to the relay (R), and $iii)$ two-way ET. We consider the problem of optimal power allocation in NC-EH-$\mathcal{RC}$ in order to maximize the total transmitted bits from S to the destination in a given time duration. General stochastic energy arrivals at S and R with known EH times and amounts are assumed. In NC-EH-$\mathcal{RC}$ with no ET, the complicated min-max optimization form along with its constraints make the problem intractable. It is shown that this problem can be transformed to a solvable convex optimization form; however, convex optimization solution does not provide the structural properties of the optimal solution. Therefore, following an alternative perspective, we investigate conditions on harvesting process of S and R where we find optimal algorithmic solution. Further, we propose some suboptimal algorithms and provide some examples, in which the algorithms are optimal. Moreover, we find a class of problems for NC-EH-$\mathcal{RC}$ with one-way ET from S to R, where the optimal algorithmic solution is devised. For NC-EH-$\mathcal{RC}$ with two-way ET, we propose \emph{general} optimal algorithmic solution. Furthermore, the performance of the proposed algorithms are evaluated numerically and compared with optimal numerical convex optimization tools.  
\end{abstract}


% \fi
\IEEEpeerreviewmaketitle

\section{Introduction} \label{sec:intro}
\IEEEPARstart{R}{ecently}, Energy Harvesting (EH) has received considerable research interest as a promising solution to the perennial energy constraint of wireless networks with limited batteries \cite{sudev}. Moreover, in near future, increasing energy consumption of highly-demanded mobile data networks is anticipated to be the main cause of global warming. Hence, EH has emerged to be used as a foundation of green communication networks \cite{zhu}. Energy harvesters collect ambient energy from the environment (including solar, hydro, wind, biomass, vibration, geothermal, piezoelectricity) and convert it into usable electrical energy. In contrast to the conventional battery-powered nodes, EH nodes have access to an unlimited source of energy which is free for users. However, the limitations in EH nodes are the low EH production rate as well as its sporadic nature. To overcome these limitations, sophisticated utilization of scavenged energy is mandatory.

Another related novel research avenue focuses on providing the power of devices wirelessly through ambient Radio Frequency (RF) signals. This avenue, known as wireless energy transfer, is motivated by notable development for the coupled magnetic resonators in \cite{Kurs} and has considerable increasingly emerging applications  \cite{Yakovlev, Kim, Huang_WET}.  
Also, recently authors in \cite{wiley} designed an efficient rectenna, which is capable of harvesting ambient RF energy. Wireless energy transfer consists of two research directions: one direction considers \emph{Simultaneous Wireless Information and Power Transfer} and characterizes the achievable rate-energy trade-off (see e.g., \cite{zhang-WIPT}, \cite{Zhou_swipt} and the references therein). Another direction aims to design a new type of networks, called \emph{Wireless Powered Communication Networks}, where the nodes harvest their required powers from wireless power transfer (see e.g., \cite{Huang_WET},\cite{Ju_WET}, and the references therein).

\subsection{Related Work and Motivation}
EH has been considered as a facility to ameliorate the energy consumption challenge of sensor nodes in many pioneering works \cite{raghun, kansal, sharma}. Information theoretic capacity of AWGN channels with an EH transmitter has been derived in \cite{ozel_TIT}. In a similar work, \cite{rajesh_TIT} has derived the shannon capacity of sensor nodes by considering processing energy cost, energy inefficiencies and channel fading. In \cite{yang}, the authors have studied the optimal packet scheduling problem in wireless single-user EH communication system, where energy and data packets are stochastically arrived at the source node: to minimize the transmit time of the data packets, transmission rate adaptively changes according to data and energy traffics. This optimal packet scheduling has later been extended to fading channel \cite{ozel_jsac}, broadcast channel \cite{antepli,yang_ozel}, multiple-access channel \cite{yang_jcn}, two-hop channel \cite{gunduz} and interference channel \cite{tutun}.

The wireless Relay Channel (RC) is a basic model to investigate the benefits of cooperation in communication networks from many aspects such as information theoretic capacity, diversity, outage analysis, cooperative and network coding, resource allocation, etc. In addition, resource-constrained networks such as Wireless Sensor Network (WSN) can get more benefit of cooperation through optimal allocation of energy and bandwidth to the nodes based on the available channel state information of those nodes (see e.g. \cite{hong} and the references therein). Motivated by the advantages that the EH and cooperation provide for the next generation wireless networks (such as high data rates, energy efficiency, and so on), a fundamental question is to find the optimal resource allocation in a RC with EH nodes.

Some special cases of multi-hop and relay channel with EH transmitting nodes have been considered in \cite{gunduz},\cite{orhan_CISS, huang}. The authors in \cite{gunduz} have considered an EH two-hop network with only the relay node harvesting the energy. In \cite{orhan_CISS}, authors have studied a two-hop network where both Source (S) and Relay (R) are the EH nodes. In \cite{huang}, \emph{half-duplex} orthogonal RC with Decode-and-Forward (DF) relay has been considered and two different delay constraints are investigated: one-block decoding delay constraint and arbitrary decoding delay constraint (up to total transmission blocks). On the other hand, Full-Duplex (FD) protocols has emerged recently to overcome the spectral efficiency loss of half-duplex protocols, by allowing the users to send and receive information concurrently at a same frequency band (see e.g., \cite{FD1},\cite{wiley_FD} and the references therein). In \cite{IWCIT}, we have considered the general model for RC with a \emph{direct link} and \emph{FD coherent} DF relaying strategy. So a more complicated min-max optimization problem has arisen in \cite{IWCIT} which has not been encountered in prior works. 
The complicated min-max problem was transformed to a solvable convex optimization form, using some mathematical background. First, an auxiliary parameter was introduced and then a minimax theorem of \cite{terk} was used to make the problem tractable. 
However, the convex optimization solutions derived in \cite{IWCIT} for FD RC do not provide detailed structural properties of optimal transmission policy. In fact, general algorithmic solution for the FD coherent DF Gaussian RC has not been tackled in the previous literatures. Moreover, this problem is not easily reducible to other channels like point-to-point, multiple-access channel, broadcast channel, two-hop channel, etc. 
None of the aforementioned works have considered the energy transfer. 
In the context of wireless powered communication networks, the authors in \cite{Gurakan} have introduced the notion of \emph{energy cooperation} where users share a portion of their scavenged energy in order to shape and optimize the energy arrivals to improve the overall performance. Here, cooperation is performed in the battery energy level instead of signal level as in the classical cooperative networks. 

\begin{figure*}[!ht]
\centering
\includegraphics[width=.6\linewidth]{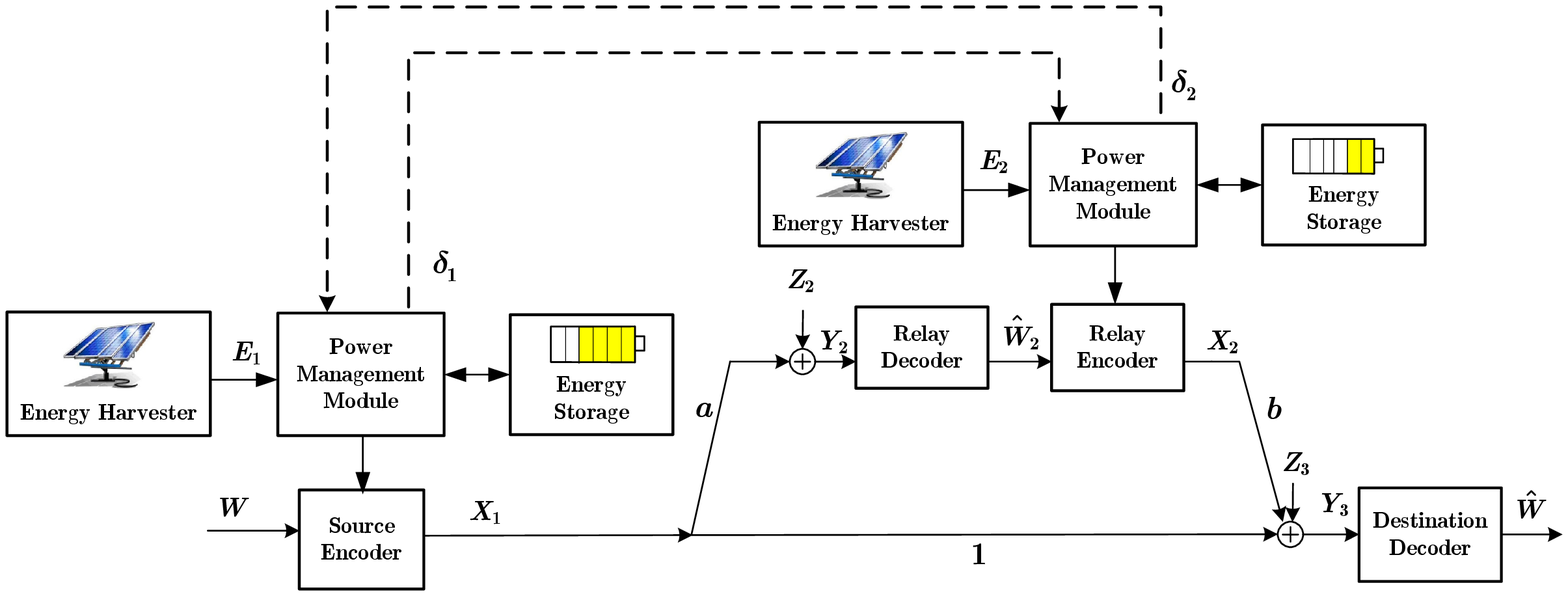}
\caption{General model for Gaussian RC with EH nodes that are capable of transferring energy to each other.}
\label{fig:1}
\end{figure*} 

\subsection{Main Contributions and Organization}
In this paper, we consider the problem of optimal power allocation for a three-node FD Gaussian RC with EH nodes. We focus on noncoherent DF relaying strategy compared to the coherent strategy in \cite{IWCIT}. Although the noncoherent DF bound on the capacity of the RC is lower than that of the coherent DF lower bound (which is the capacity of the degraded RC); implementing noncoherent communication is more convenient in wireless systems. 
Our goal is to maximize the total number of bits that can be delivered from S node to the destination (D) node in a given time duration. Three cases are studied based on the ability of the nodes to transfer some parts of their harvested energy: $(i)$ no Energy Transfer (ET) among nodes $(ii)$ one-way ET from S to R, and $(iii)$ two-way ET between S and R or bi-directional energy cooperation. 
We consider a general model compared to the existing works. In our model, there is a \emph{direct link} from S to D (in contrast to \cite{gunduz},\cite{orhan_CISS}, \cite{Gurakan}) and also we investigate the FD mode compared to the half-duplex mode of \cite{huang}. Besides, unlike \cite{gunduz},\cite{orhan_CISS, huang},\cite{IWCIT}, we consider the energy transfer among nodes, which makes our model more general. We assume zero cost energy transfer among nodes. Studying the cost of energy sharing among nodes is parallel to our work (see e.g., \cite{Gurakan} and the references therein).  
In this work, we investigate the offline problem where we assume the availability of offline knowledge about EH times and amounts in S and R. This is due to the fact that the online problem that assigns the nodes' powers in real-time is intractable for now in our studied model and it is consistent with the assumptions in existing works, such as \cite{yang, ozel_jsac, antepli, yang_ozel, yang_jcn, gunduz, tutun}.

In our problem, like \cite{IWCIT}, the structural properties of optimal policy can not be derived from the convex optimization solution. Therefore, we follow a different perspective to derive the algorithmic solutions. 
Our main contributions in the rest of this paper are organized as follows.

\begin{itemize}
\item In Section~\ref{sec:sysmodel}, we propose a general model for FD Non-Coherent DF Gaussian RC with EH nodes, called ``NC-EH-$\mathcal{RC}$'' in three cases: $1)$ no-ET, $2)$ one-way ET from S to R, and $3)$ two-way ET. Also, relaying strategy and harvesting process are described and some preliminaries are added to make the paper self-contained.

\item Section~\ref{sec:formulation} considers NC-EH-$\mathcal{RC}$ with no ET case. First, we formulate the power allocation problem. Then, we show that it can be transformed to a tractable form even though it is a complicated optimization problem. However, the solution do not provide the detailed structural properties of optimal solution. Hence, we explore some conditions on the harvesting process of R that help us to find the optimal algorithmic solution. We provide this solution when the R is in good EH condition, which means that R can forward any received information from S toward D without any energy shortage. This solution reveals some important specifications of general optimal solution for our problem that discriminate it from other problems solved in the literatures. Moreover it is shown that disjoint optimization at S and R is suboptimal, though it is optimal in some special cases. We further propose a suboptimal algorithmic solution based on \emph{total} power allocation for S and R, which is optimum for some realization of EH process at S and R.

\item In Section~\ref{sec: one-way}, we study optimal power allocation in NC-EH-$\mathcal{RC}$ with one-way ET from S to R. For some conditions on the harvesting process of S, we propose an algorithmic optimal solution for our power allocation problem. We devise an algorithm for optimal power allocation, when the S is in good EH condition. It means that for a fixed amount of network's energy resources, S transfers some parts of its harvested energy (stored in its battery) to R in order to improve the performance. 

\item In Section~\ref{sec: two-way}, we concentrate on the optimal power allocation for NC-EH-$\mathcal{RC}$ with two-way ET or bi-directional energy cooperation. We propose a \emph{general} algorithmic optimal solution for the problem in this case. In fact, two-way ET capability provides new interesting specifications for optimal solution. These are utilized for devising algorithms that solve the problem optimally. 
 
\item In Section~\ref{numerical results}, we evaluate the performance of our proposed algorithmic solutions derived in sections~\ref{sec:formulation}, \ref{sec: one-way}, and \ref{sec: two-way}, numerically. Also, we present some typical examples, where each of the suboptimal solutions outperforms the other one and is optimal. Besides, allocated powers of nodes in our algorithms are compared with optimal numerical convex optimization tool. 

\item Finally, section~\ref{sec:coclusion} concludes the paper.

\end{itemize} 

%In a near-optimal solution, by introducing a new constraint on the total transmission powers of the network nodes, we examine a relaxed problem with the solution that approximates the solution of the main problem. In the next, we propose an algorithmic solution that meets the constraints of the main problem and hence provide a suboptimal algorithmic solution of the main problem.

%The ~ before \ref is to prevent a line break just at this position.
\section{System Model and Preliminaries}\label{sec:sysmodel}
\textbf{Notation}: Upper-case letters (e.g., $X$) denote Random Variables (RVs) and lower-case letters (e.g., $x$) their realizations. The probability mass function (p.m.f) of a RV $X$ with alphabet set $\mathcal{X}$ is denoted by $p_X(x)$.
%occasionally subscript $X$ is omitted.
%$A_\epsilon^n(X,Y)$ is the set of $\epsilon$-strongly, jointly typical sequences of length $n$.
%$X^j_i$ indicates a sequence of RVs $(X_i,X_{i+1},...,X_j)$; we use $X^j$ instead of $X^j_1$ for brevity. 
The variables related to S and R are indicated with subscripts $1$ and $2$, respectively. We show $N$-length vectors by bold-face letters, (e.g.,$\mathbf{V_i}=[V_i^1,\,V_i^2,\,...,V_i^N]$), where their $j$-th component  is denoted by superscript $j$ (e.g., $V_i^j$). 
$\mathcal {CN}(0,\sigma^2)$ denotes a zero-mean complex value Gaussian distribution with variance $\sigma^2$ and $I(X;Y)$ denotes the \emph{mutual information} between $X$ and $Y$. In addition, $\mathcal{C}(x)={\log_2}(1+x)$ and $[x]^\dag=\max \left\{ {1,x} \right\}$.
%$\|\mathbf{X}\|_p$ is the $L^p$-norm of a vector $\mathbf{X}$; $\mathbf{X}(i)$ is its $i$th element. $(\cdot)^T$, $(\cdot)^\dag$ and $\Nc(\cdot)$ denote the transpose, conjugate transpose and null space operations, respectively. For stating asymptotic results (Landau notation), $f(n)=o(g(n))$ if $\lim\limits_{n\rightarrow \infty}\frac{f(n)}{g(n)}\rightarrow 0$.

RC models a three-node network, in which the source node wants to communicate to the destination node with the help of the relay node.
Codebook, encoder, decoder and rate for the discrete memoryless RC (DM-RC) can be defined as \cite[Chapter~16]{El_Gamal}.

A general model for Gaussian RC with EH nodes and ET capabilities is depicted in Fig.~ \ref{fig:1}. 
The channel outputs corresponding to the channel inputs $ {X_1},{X_2} $ are as follows
\begin{IEEEeqnarray}{rCl}
{Y_2} & = & a{X_1} + {Z_2},\label{y_relay}\\
{Y_3} & = & {X_1} + b{X_2} + {Z_3},\label{y_destination}
\end{IEEEeqnarray}
where $ a $ and $ b $ are channel gains of S-R and R-D links, respectively, assuming normalized channel gain for the S-D link, and we have $ {Z_2} \sim {\cal CN}(0,N_0),\,\,{Z_3} \sim {\cal CN}(0,N_0) $.

The energy harvester block scavenges the ambient energy from the environment. Transmitting nodes are capable of transferring parts of their harvested energies to each other to have better control on the network's energy resources. We study three cases in this paper, namely $(i)$ nodes with no ET ($\mathbf{\delta_1}=\mathbf{\delta_2}=0$ in Fig.~\ref{fig:1}), $(ii)$ nodes with one-way ET from S to R ($\mathbf{\delta_1}\neq 0,\,\mathbf{\delta_2}=0$ in Fig.~\ref{fig:1}), and $(iii)$ nodes with two-way ET ($\mathbf{\delta_1}\neq 0,\,\mathbf{\delta_2} \neq 0$ in Fig.~\ref{fig:1}), where $\mathbf{\delta_1}$ and $\mathbf{\delta_2}$ denote the amount of energy transferred in S $\rightarrow$ R and R $\rightarrow$ S directions, respectively. The power management module determines the amount of the harvested energy used for communication as well as the amount transferred to other node. The remaining energy is stored in the energy storage device (e.g., battery or super-capacitor) for future use. 

\subsection{Relaying Strategy}

Since the capacity of FD DM-RC is not known in general, in this paper, we consider an achievable rate for the RC that provides a lower bound on its capacity. This rate is achieved by \emph{noncoherent} DF strategy in the R. Since implementing of the coherent communication is difficult in wireless systems \cite{El_Gamal}, it is more convenient to consider noncoherent coding schemes in S and R at the cost of loosing some rate. In this strategy, R plays a central role in communication by decoding the information bits received from S. Then, S and R noncoherently cooperate to transmit their codewards to D. D decodes the received coded information from S and R, simultaneously. With this coding scheme, the following rate for DM-RC is achievable \cite[Chapter~16]{El_Gamal}:

\begin{equation}
C \ge \mathop {\max }\limits_{p_{X_1}({x_1})p_{X_2}({x_2})} \min \left\{ {I({X_1},{X_2};{Y_3}),I({X_1};{Y_2}|{X_2})} \right\}. \label{N-DF lower bound}
\end{equation} 

Note that $X_1$ and $X_2$ are independent in this case. Hence, we used $p_{X_1,X_2}({x_1},{x_2})=p_{X_1}({x_1})p_{X_2}({x_2})$.  The corresponding rate for the Gaussian RC is given by
\\

$C \ge \min \left\{ {\mathcal{C}\left( {\frac{{{P_1} + {b^2}{P_2}}}{N_0}} \right),\mathcal{C}\left( {\frac{{\max \{ 1,{a^2}\} {P_1}}}{N_0}} \right)} \right\}$
\begin{equation}
\,\,\,\,\,\,=\left\{
\begin{IEEEeqnarraybox}[\IEEEeqnarraystrutmode
\IEEEeqnarraystrutsizeadd{2pt}{2pt}][c]{lll}
{\mathcal{C}\left( {\frac{{{P_1} + {b^2}{P_2}}}{N_0}} \right)}   \qquad\qquad \textrm{if} \;\;\; {\frac{{({[a^2]^\dag} - 1){P_1}}}{{b^2{P_2}}} > 1},\\
{\mathcal{C}\left( {\frac{{[a^2]^\dag {P_1}}}{N_0}} \right),} \qquad\qquad\;\;\;  \textrm{otherwise},
\end{IEEEeqnarraybox}
\right.\label{noncoherent}
\end{equation}
where $P_1$ and $P_2$ are the powers of S and R, respectively. The first term under the minimum can be interpreted as \emph{Noncooperative Multiple-Access} (N-MAC) term. The second term implies that if the quality of S-R link is worse than that of the direct S-D link (i.e., $a^2<1$), any information that R can decode, is previously decoded at D. In such cases, R cannot help and should be ignored (The minimum is ${\mathcal{C}\left( {\frac{{{P_1}}}{N_0}} \right)}$, which is the capacity of direct link from S to D). We do not consider these cases in this paper. 
% and the second is the C-BC bound.

\begin{remark}
In two-hop networks, studied in \cite{orhan_CISS, orhan_sarnoff}, there is a \emph{data causality constraint} at R: data bits can only be transmitted from R toward D after they have arrived from S. With this constraint, in a given time duration, the minimum of total bits transmitted from S to R (called $\mathcal{B}_{S\rightarrow R}$) and the total transmitteed bits from R to D (called $\mathcal{B}_{R\rightarrow D}$) is equal to $\mathcal{B}_{R\rightarrow D}$. Therefore, the min-max problem has not been encountered in these works.  
This again shows the complexity of our problem compared to that of \cite{orhan_CISS, orhan_sarnoff}.   
\end{remark}

\subsection{Harvesting Process}
\begin{figure}[tb]
\centering
\includegraphics[width=1\linewidth]{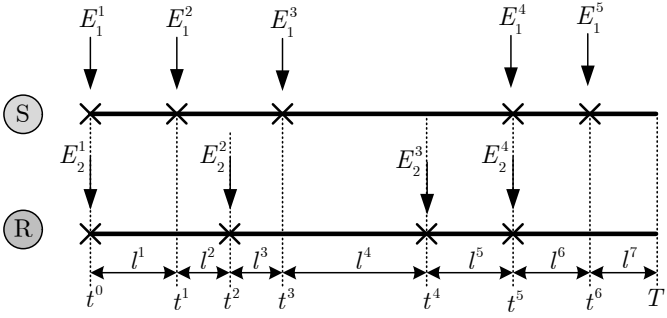}
\caption{EH instants and amounts for S and R with $ K=6 $.}
\label{fig:2}
\end{figure}

We consider a RC, in which the harvested energy from the environment is the sole source of energy in the network. 
Our problem is to maximize the number of bits delivered by a deadline $ T $ from S to D. S and R harvest energy at random instants $ {t^0},{t^1},{t^2},...,{t^K} $ and in random amounts $ E_1^1,E_1^2,...,E_1^{K+1} $ and $ E_2^1,E_2^2,...,E_2^{K+1} $, respectively. If at some instants only S or R harvests energy, we simply set the amounts of the energy harvested by the other one to zero (see Fig. \ref{fig:2}).
The interval between two harvesting instants is called an epoch. The length of $ i^{th} $  epoch is  $ {l^i} = {t^i} - {t^{i - 1}} $ for $ i = 1,...,K+1 $. So, there are a total of $ K+1 $ epoch with $ {t^0}=0 $ and $ {l^{K + 1}} = T - {t^K} $. We consider the offline problem in which the ${t^i},{E_1^i},{E_2^i},\,\forall i$ are known to S and R before the start of transmission. Moreover, considering the offline problem, enables transmitting nodes to share their harvested energies in the offline mode. We ignore any inefficiency in ET among nodes. 

\subsection{Optimal Point-to-Point Solution and Its Interpretation}

\begin{figure}[tb]
\centering
\includegraphics[width=1\linewidth]{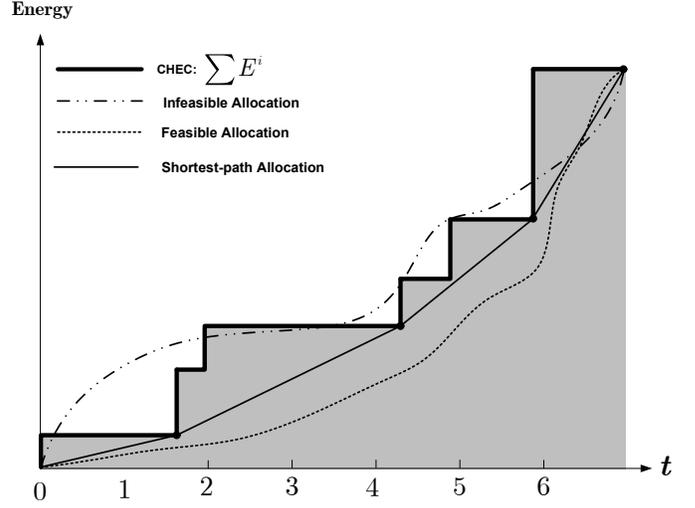}
\caption{Shortest path power allocation for point-to-point EH communication system. A typical feasible and infeasible solutions are included in this figure for a typical cumulative energy harvested curve (CEHC), shown in thick solid line.}
\label{shortest_path}
\end{figure}

To make the paper self-contained, we briefly describe the optimal packet scheduling in wireless point-to-point EH communication systems which was proposed in \cite{yang}, translating to our notations. The problem in \cite{yang} is to find the optimal offline transmission policy, which minimizes the transmission completion time with a predefined data bits to transmit. Energies harvested at time instants $t^i$ with the amount of $E^i$ at the transmitter. This problem is shown to be the dual of maximizing the throughput (total number of bits that can be transmitted) in a given time (deadline), with the same optimal transmission policies \cite{ozel_jsac}. The solution for these problems is as follows \cite[Theorem 1]{yang}:
%\begin{equation}
%{o_n} = \mathop {\argmin }\limits_{{o_{n - 1}} < i \le {K+1}} \frac{{\sum\nolimits_{j = {o_{n - 1}}}^{i-1} {E^j} }}{{{t^i} - {t^{{o_{n - 1}}}}}},
%\end{equation}
%
%
%\begin{equation}
%{ {P^n}}^* = \frac{{\sum\nolimits_{j = {o_{n - 1}}}^{{o_n} - 1} { E^j} }}{{{t^{{o_n}}} - {t^{{o_{n - 1}}}}}},
%\end{equation}
%
%\begin{equation}
%{ {l^n}}^* = {{t^{{o_n}}} - {t^{{o_{n - 1}}}}},
%\end{equation}
%where $T=\sum_{n=1}^{N}{ {l^n}}^*$ and ${o_n}$ is the index of time instant that allocated power changes from $P^n$ to $P^{n+1}$.
\begin{equation}
{i_n} = \mathop {\argmin }\limits_{{i: t^{i_n}-1 < {t^i}} \le T} \frac{{\sum\nolimits_{j = {i_{n - 1}}}^{i-1} {E^j} }}{{{t^i} - {t^{{i_{n - 1}}}}}},\;n=1,...,N
\end{equation}

\begin{equation}
{ {P^n}} = \frac{{\sum\nolimits_{j = {i_{n - 1}}}^{{i_n} - 1} { E^j} }}{{{t^{{i_n}}} - {t^{{i_{n - 1}}}}}},\;n=1,...,N
\end{equation}

\begin{equation}
{ {L^n}} = {{t^{{i_n}}} - {t^{{i_{n - 1}}}}}=\sum_{j=i_{n-1}+1}^{i_n}{ {l^j}},\;n=1,...,N
\end{equation}
where $P^i,\,\forall i$ is the sequence of transmission power with corresponding duration of $L^i,\,\forall i$, respectively, $T=\sum_{n=1}^{N}{ {L^n}}$ and ${i_n}$ is the index of time instant that allocated power changes from $P^n$ to $P^{n+1}$.
This result is achieved using some lemmas about necessary optimality conditions. The idea follows from the lazy scheduling \cite{uysal2002}, convexity \cite{boyd} and majorization theory \cite{Majorization}. 
The solution has the \emph{shortest-path} graphical interpretation. This notion is depicted in Fig.~\ref{shortest_path}: any power allocation restricted to the Cumulative Energy Harvested Curve (CEHC), $\sum{ {E^i}}$, from below is feasible; It is infeasible otherwise. The optimal solution is shown to be the piecewise  linear curve with the shortest length (or tightest string) restricted to CEHC from below, connecting the origin to the end point of CEHC (shown by thin solid line in Fig.~\ref{shortest_path}).

In next three sections, we investigate the optimal algorithmic power allocation solution in our NC-EH-$\mathcal{RC}$ model with no ET, one-way ET and two-way ET. 

%-----------------------------------------------------
\newcounter{tempequationcounter}
\begin{figure*}[!t]
\normalsize
\setcounter{tempequationcounter}{\value{equation}}
\begin{IEEEeqnarray*}{rCl}
\setcounter{equation}{25}
{\cal L}(\{ P_1^i\} ,\{ P_2^i\} ,\xi ,\mu ,\vartheta ,\eta ) &=& \sum\limits_{i = 1}^{K + 1} {\left\{ {{\lambda ^i}\mathcal{C}\left( {\frac{{{P_1^i} + {b^2}{P_2^i}}}{N_0}} \right) + (1 - {\lambda ^i})\mathcal{C}\left( {\frac{{[a^2]^\dag P_1^i}}{N_0}} \right)} \right\}\,{l^i}}\\
&-& \sum\limits_{k = 1}^{K + 1} {{\xi _k}\left( {\sum\limits_{i = 1}^k {P_1^i{l^i}}  - \sum\limits_{i = 0}^{k - 1} {E_1^i} } \right)}  - \sum\limits_{k = 1}^{K + 1} {{\mu _k}\left( {\sum\limits_{i = 1}^k {P_2^i{l^i}}  - \sum\limits_{i = 0}^{k - 1} {E_2^i} } \right)}  + \sum\limits_{i = 1}^{K + 1} {{\vartheta _i}P_1^i}  + \sum\limits_{i = 1}^{K + 1} {{\eta _i}P_2^i} ,\yesnumber\label{longequation}
\end{IEEEeqnarray*}
\setcounter{equation}{\value{tempequationcounter}}
\hrulefill
\vspace{4pt}
\end{figure*}
%---------------------------------------------------------

\section{NC-EH-$\mathcal{RC}$ with No Energy Transfer}\label{sec:formulation}

In this section, we consider the NC-EH-$\mathcal{RC}$ with no ET. We study the optimal power allocation for S and R in order to maximize the total transmitted bits from S to D, satisfying energy causality constraints at S and R. This means that energy cannot be utilized in S and R before it is harvested in the corresponding node. We consider the noncoherent relaying strategy, achieving the rate in (\ref{noncoherent}). Therefore, we can formulate the problem as:

\begin{equation}
\!\!\mathop {\max }\limits_{{P_1},{P_2}} \sum\limits_{i = 1}^{K+1} {\min } \left\{ {\tilde C_1^{}(P_1^i,P_2^i),\tilde C_2^{}(P_1^i)} \right\}{l^i}
%\! = \!\mathop {\max }\limits_{{P_1},{P_2}} \sum\limits_{i = 1}^{K+1} {C(P_1^i,P_2^i)} 
\label{main-no et}
\end{equation}
\begin{IEEEeqnarray}{rll}
s.t.\qquad& P_1^i \ge 0,\,P_2^i \ge 0,&\qquad i = 1,...,K + 1,\label{nonneg}\\
& \sum\limits_{i = 1}^k {P_1^i{l^i} \le \sum\limits_{i = 0}^{k - 1} {E_1^i} } ,&\qquad k = 1,...,K + 1,\label{Scausal}\\
&\sum\limits_{i = 1}^k {P_2^i{l^i} \le \sum\limits_{i = 0}^{k - 1} {E_2^i} } ,&\qquad k = 1,...,K + 1.\label{Rcausal}
\end{IEEEeqnarray}
where we have $\tilde C_1^{}(P_1^i,P_2^i)=\mathcal{C}\left( {\frac{{{P_1^i} + {b^2}{P_2^i}}}{N_0}} \right)$ and $\tilde C_2^{}(P_1^i)=\mathcal{C}\left( {\frac{{[a^2]^\dag {P_1^i}}}{N_0}} \right)$.

Equation (\ref{nonneg}) denotes the non-negativity of the powers at S and R. Equations (\ref{Scausal}) and (\ref{Rcausal}) state the energy causalities at S and R, respectively. 
Finding the solution of the above problem is not straightforward as it has the min-max optimization form which cannot be separated due to the FD nature of the problem. In other words, since R sends and receives information at the same time, in each epoch we do not know which term (in epoch $ i $, $\tilde C_1^{}(P_1^i,P_2^i) $ or $ \tilde C_2^{}(P_1^i) $)  is the minimum. Therefore, optimal power assignment for S and R to maximize the total transmitted bits is unknown. Also, observe that in (\ref{noncoherent}) the condition that specifies the minimum term depends on the optimization parameters (i.e., $ {\mathbf{P_1}} $ and $ {\mathbf{P_2}} $ in (\ref{main-no et})-(\ref{Rcausal})).

%\section{Optimal Solution for FD Gaussian RC}\label{sec:optimalsol}
%In \cite{IWCIT}, we proposed a technique to make the problem in (\ref{main-no et})-(\ref{Rcausal}) tractable. That technique is applied here as our cost function in (\ref{noncoherent}) has similar form to the one in \cite{IWCIT}. 
We rewrite the problem in (\ref{main-no et})-(\ref{Rcausal}) by introducing $ 0 \le \lambda^i  \le 1 $ as follows
\begin{equation}
\mathop {\max }\limits_{\{ P_{_1}^i\} ,\{ P_2^i\}} \sum\limits_{i = 1}^{K+1} {\mathop {\min }\limits_{\{ {\lambda ^i}\} }  } \,\left\{ {{\lambda ^i}\tilde C_1^{}(P_1^i,P_2^i) + (1 - {\lambda ^i})\tilde C_2^{}(P_1^i)} \right\}{l^i}\label{opt_prob}
\end{equation}
\begin{IEEEeqnarray}{rll}
s.t. \qquad& P_1^i \ge 0,\,P_2^i \ge 0,& \qquad i = 1,...,K + 1,\\
&\sum\limits_{i = 1}^k {P_1^i{l^i} \le \sum\limits_{i = 0}^{k - 1} {E_1^i} },&\qquad k = 1,...,K + 1,\\
&\sum\limits_{i = 1}^k {P_1^i{l^i} \le \sum\limits_{i = 0}^{k - 1} {E_1^i} },&\qquad k = 1,...,K + 1,\label{opt_prob_end}
\end{IEEEeqnarray}
where,
\begin{equation}
{\lambda ^i} = 
\begin{cases}
0& \textrm{if}\qquad \tilde C_1^{}(P_1^i,P_2^i) > \tilde C_2^{}(P_1^i),\\
1&\textrm{if}\qquad \tilde C_1^{}(P_1^i,P_2^i) < \tilde C_2^{}(P_1^i),\\
\textrm{arbitrary}& \textrm{if}\qquad \tilde C_1^{}(P_1^i,P_2^i) = \tilde C_2^{}(P_1^i).
\end{cases}
\end{equation}

Noting the similarity of the problem in (\ref{opt_prob}) with the one in \cite{IWCIT}, we use the technique proposed in {\cite[Theorem 1]{IWCIT}} and change the order of min and max operators in (\ref{opt_prob}) by applying the min-max theorem of Terkelsen \cite{terk}.

Then, the problem in (\ref{opt_prob})-(\ref{opt_prob_end}) can be decomposed into the following two problems.

\begin{equation}
\begin{array}{l}
{\!\!\!\!\!({\rm{Problem}}\,{\rm{1}}):{f^*}(\{ {\lambda ^i}\} )\! = \!\mathop {\max }\limits_{\{ P_{_1}^i\} ,\{ P_2^i\} } \sum\limits_{i = 1}^{K + 1}\! {\left\{ {{\lambda ^i}\mathcal{C}\left( {\frac{{{P_1^i} + {b^2}{P_2^i}}}{N_0}} \right)} \right.}} \\{
\,\,\,\,\,\,\,\,\,\,\,\;\;\;\;\,\,\,\qquad\qquad\,\,\,\,\,\,\,\,\,\,\,\,\,\,\,\,\,\,\,\,\,\,\,\,\,\,\,\,\,\,\,\,\,\,\,\,\,\,\,\left. { + (1 - {\lambda ^i})\mathcal{C}\left( {\frac{{[a^2]^\dag {P_1^i}}}{N_0}} \right)} \right\}}{l^i}
\end{array}
\end{equation}
\begin{IEEEeqnarray}{rll}
s.t.\qquad& P_1^i \ge 0,\,P_2^i \ge 0,&\qquad i = 1,...,K + 1,\\
&\sum\limits_{i = 1}^k {P_1^i{l^i} \le \sum\limits_{i = 0}^{k - 1} {E_1^i} },&\qquad k = 1,...,K + 1,\\
&\sum\limits_{i = 1}^k {P_2^i{l^i} \le \sum\limits_{i = 0}^{k - 1} {E_2^i} },&\qquad k = 1,...,K + 1.
\end{IEEEeqnarray}
\begin{IEEEeqnarray}{lll}
\!\!\!\!\!\!\!\!\!\!\!\!\!\!({\rm{Problem}}\,2):&\,\,\mathop {\min }\limits_{\{ {\lambda ^i}\} } \,\,\,{f^*}(\{ {\lambda ^i}\} )& \label{problem2}\\
& s.t.\,\,\,\,\,\,\,\,0 \le {\lambda ^i} \le 1,\,\,\,\,\,\,\,\,\,\,i = 1,...,K + 1.&
\end{IEEEeqnarray}
Problems 1 and 2 are convex optimization problems as their objective functions are concave and their constraints are affine; thus, they can be solved by efficient convex optimization methods \cite{boyd}. 
%that can be solved efficiently, too.

For Problem 1, we write the Lagrangian function for any $ {\xi _k} \ge 0,\,\,{\mu _k} \ge 0,{\vartheta _k} \ge 0$ and ${\eta _k} \ge 0 $ as (\ref{longequation}), in top of this page. KKT optimality conditions for (\ref{longequation}) are

\begin{IEEEeqnarray}{lll}
\frac{{{\lambda ^i}}}{{P_1^i + {b^2}P_2^i}+N_0} + \frac{{(1 - {\lambda ^i}){[a^2]^\dag}}}{{ {[a^2]^\dag}P_1^i}+N_0} - \sum\limits_{k = i}^K {{\xi _k}}  +\vartheta _i &\,=\,& 0,\;\forall i\\
\frac{{  {b^2}{\lambda ^i}}}{{ P_1^i + {b^2}P_2^i}+N_0} - \sum\limits_{k = i}^K {{\mu _k}}+\eta _i &\,=\,& 0,\;\forall i
\end{IEEEeqnarray}
together with following complementary slackness conditions;
\addtocounter{equation}{1}
\begin{IEEEeqnarray}{rl}
{\xi _k}\left( {\sum\limits_{i = 1}^k {P_1^i{l^i}}  - \sum\limits_{i = 0}^{k - 1} {E_1^i} } \right) = 0, &\;\;\; k = 1,...,K, \label{slackness1}\\
{\mu _k}\left( {\sum\limits_{i = 1}^k {P_2^i{l^i}}  - \sum\limits_{i = 0}^{k - 1} {E_2^i} } \right) = 0,&\;\;\; k = 1,...,K,\label{slackness2}\\
\sum\limits_{i = 1}^N {{\vartheta _i}P_1^i}  = 0,&\;\;\; i = 1,...,K + 1,\label{slackness3}\\
\sum\limits_{i = 1}^N {{\eta _i}P_2^i}  = 0,&\;\;\; i = 1,...,K + 1.\label{slackness4}
\end{IEEEeqnarray}

We find the optimal solution as

\begin{equation}
P_1^{i*} = \frac{{(1 - {\lambda ^i})}}{{\frac{{ - 1}}{{{b^2}}}\sum\limits_{k = i}^K {{\mu _k}}  + \sum\limits_{k = i}^K {{\xi _k}}+\frac{\eta_i}{b^2}-\vartheta _i }} - \frac{N_0}{{{[a^2]^\dag}}},\,\,\,\,\,\forall i \label{p1-star}
\end{equation}

\begin{equation}
P_2^{i*} = \frac{{{\lambda ^i}}}{{\sum\limits_{k = i}^K {{\mu _k}}-\eta_i }} - \frac{{(1 - {\lambda ^i})}}{{{b^2}\sum\limits_{k = i}^K {{\xi _k}}  - \sum\limits_{k = i}^K {{\mu _k}}-b^2\vartheta _i+{\eta_i} }} + \frac{N_0}{{{[a^2]^\dag}{b^2}}} - \frac{N_0}{{{b^2}}},\,\,\,\forall i. \label{p2-star}
\end{equation}

To be able to solve Problem 2, one has to find the $P_1^{i*}$ and $P_2^{i*}$ in terms of only ${\lambda ^i}$. Then, it is enough to solve the following problem with respect to ${\lambda ^i}$s.

\begin{IEEEeqnarray}{l}
\!\!\!\!\!\!\!\!\mathop {\min }\limits_{\{ {\lambda ^i}\} } \sum\limits_{i = 1}^{K + 1} {\left( {{\lambda ^i}\mathcal{C}(\frac{P_1^{i*} + {b^2}P_2^{i*}}{N_0}) + (1 - {\lambda ^i})\mathcal{C}(\frac{{[a^2]^\dag}P_1^{i*}}{N_0})} \right)}{l^i}  \\
 s.t.\,\,\,\,\,\,\,\, 0 \le {\lambda ^i} \le 1,\,\,\,\,\,\,\,\,\,\,i = 1,...,K + 1.
\end{IEEEeqnarray}

%\begin{remark}
%As Problem 2 is linear in terms of ${\lambda ^i}$, its extremum point occurs at the corner points. This means that we have ${{\lambda ^i}^*}\in \{0,1\}, \forall i$   
%\end{remark}

As the solutions provided in (\ref{p1-star}) and (\ref{p2-star}) do not give any explicit idea about the structural properties of optimal power assignment; therefore, it is not straightforward to find algorithmic solutions for $\mathbf{P_1^{*}}$ and $\mathbf{P_2^{*}}$ using these expressions. 
In fact, finding a general algorithmic solution for optimal power allocation for NC-EH-$\mathcal{RC}$ is a complex and non-trivial task that has not been tackled in the existing works, yet.
Therefore, in the following we find the optimal algorithmic solution for NC-EH-$\mathcal{RC}$ with no ET in a special case to gain insight on the optimal solution. 

\subsection{Optimal Algorithmic Solution for NC-EH-$\mathcal{RC}$ with No ET}
In this case, the R is in good EH condition and has scavenged sufficient energy such that it is able to forward any information bits received from S.
Algorithm \ref{algorithm special 1} gives the optimal solution in this case. Its optimality is shown in Lemma~\ref{lem alg R good}.

\begin{algorithm} 
\caption{Optimal greedy power allocation algorithm for NC-EH-$\mathcal{RC}$ with no ET, when R is in good EH condition.}
\label{algorithm special 1}
\begin{algorithmic}
\STATE (1) \emph{Single-user Power Allocation for S}

    ${o_n} = \mathop {\argmin }\limits_{{o_{n - 1}} < i \le {K+1}} \frac{{\sum\nolimits_{j = {o_{n - 1}}}^{i-1} {E_1^j} }}{{{t^i} - {t^{{o_{n - 1}}}}}}$

    ${P_1^n}^* = \frac{{\sum\nolimits_{j = {o_{n - 1}}}^{{o_n} - 1} {E_1^j} }}{{{t^{{o_n}}} - {t^{{o_{n - 1}}}}}}$

\STATE (2) \emph{Feasibility Problem for Power Allocation at R}
    \IF {$P_2^i=\frac{[a^2]^\dag-1}{b^2}{P_1^i}^*,\,\forall i$ is feasible}
        \STATE Find power allocation for R as ${P_2^i}^*=\frac{[a^2]^\dag-1}{b^2}{P_1^i}^*,\,\forall i$ 

    \ELSE
        \RETURN The algorithmic optimal solution is not known in general.
    \ENDIF 

\end{algorithmic}
\end{algorithm}

\begin{defi}
We define the class of problem (\ref{noncoherent}) with \emph{R in good EH condition} as the problems, where $P_1^{i*},\,P_2^{i*},\,\forall i$ (in optimal solution) satisfy
\begin{align*}
{P_2^i}^*\ge \frac{[a^2]^\dag-1}{b^2}{P_1^i}^*,\,\forall i.
\end{align*}
This means that R harvests sufficient energy so that at any time instant, it has enough power to transfer any bits received from S toward D. The algorithm is called \emph{greedy} when R uses the least power to transmit the data bits received by S, i.e., ${P_2^i}^*= \frac{[a^2]^\dag-1}{b^2}{P_1^i}^*,\,\forall i$.  
\end{defi}

\begin{lemma}\label{lem alg R good}
In NC-EH-$\mathcal{RC}$ with no ET, Algorithm \ref{algorithm special 1} provides the optimal greedy power allocation, when R is in good EH condition.
\end{lemma}

\begin{IEEEproof}
If R is in good EH condition, the bottleneck is the S-R link. Hence, the cost function is expressed as the second term under the minimum and the problem is

\begin{IEEEeqnarray*}{l}
\!\!\mathop {\max }\limits_{{P_1},{P_2}} \sum\limits_{i = 1}^{K+1}{\mathcal{C}\left( {\frac{{{[a^2]^\dag} {P_1^i}}}{N_0}} \right)}{l^i} \\
s.t. \qquad (\ref{nonneg})-(\ref{Rcausal})
\end{IEEEeqnarray*}

Removing variables and constraints irrelevant to the cost function results in

\begin{IEEEeqnarray*}{l}
\!\!\mathop {\max }\limits_{{P_1}} \sum\limits_{i = 1}^{K+1}{\mathcal{C}\left( {\frac{{{[a^2]^\dag} {P_1^i}}}{N_0}} \right)}{l^i} \\
s.t. \qquad P_1^i \ge 0,\; \forall i\;\;\; \mathrm{and} \;(\ref{Scausal})
\end{IEEEeqnarray*}

Now, the cost function is a concave function of its single variable $\mathbf{P_1}$ and the constraints are convex sets over $\mathbf{P_1}$. Hence, the problem is convex and the solution is the shortest-path power allocation algorithm for the S. Now, R should only use sufficient power to make $\tilde C_1^{}(P_1^i,P_2^i)\ge \tilde C_2^{}(P_1^i),\, \forall i$ or ${ \mathcal{C}\left(\frac{P_1^i+b^2 P_2^i}{N_0}\right)\ge \mathcal{C}\left( {\frac{{{[a^2]^\dag} {P_1^i}}}{N_0}} \right)},\, \forall i$. This is equivalent to satisfy ${P_1^i+b^2 P_2^i} \ge  {{{[a^2]^\dag} {P_1^i}}},\, \forall i$, as $\mathcal{C}(.)$ is a monotonic function. Algorithm \ref{algorithm special 1} utilizes the least possible power for R in each epoch to achieve the above result, so it is called the \emph{greedy} algorithm.
\end{IEEEproof}

\begin{remark}
In existing works, the EH nodes must use all their harvested energy in order to be optimal \cite{yang, ozel_jsac, antepli, yang_ozel, yang_jcn, gunduz}, \cite{orhan_CISS}, \cite{orhan_sarnoff}. However, Lemma~\ref{lem alg R good} shows that for NC-EH-$\mathcal{RC}$, leaving some parts of harvested energy unused in the battery of R is not necessarily suboptimal. This fact shows that our problem can not be reduced to the existing EH problems in \cite{yang, ozel_jsac, antepli, yang_ozel, yang_jcn, gunduz}, \cite{orhan_CISS}, \cite{orhan_sarnoff}.
In addition, if we do not use a greedy algorithm, any feasible power allocation for R, greater than the one in Algorithm \ref{algorithm special 1}, also provides the optimal solution for the problem. Therefore, Lemma~\ref{lem alg R good} reveals a general specification of the solution of our general problem in (\ref{main-no et})-(\ref{Rcausal}) in the following lemma. 
\end{remark}

\begin{lemma}
The optimal power allocation for NC-EH-$\mathcal{RC}$ is not necessarily unique.
\end{lemma} 

\begin{IEEEproof}[Outline of proof]
Consider an optimal power allocation that in each epoch the second term under the minimum of (\ref{noncoherent}) is dominant (the achievable rate of channel is forced by the rate of S-R channel). In such cases, it is obvious that expending more power by the R, not violating its energy causality constraint, has no benefit in terms of total transmitted bits in a given deadline and leads to the same optimal value. Therefore, $\mathbf{P_1^{*}}$ and $\mathbf{P_2^{*}}$ are not necessarily unique. 
\end{IEEEproof}

Now, we present an example that applies to the considered case.

\begin{figure}[tb]
\centering
\includegraphics[width=1\linewidth]{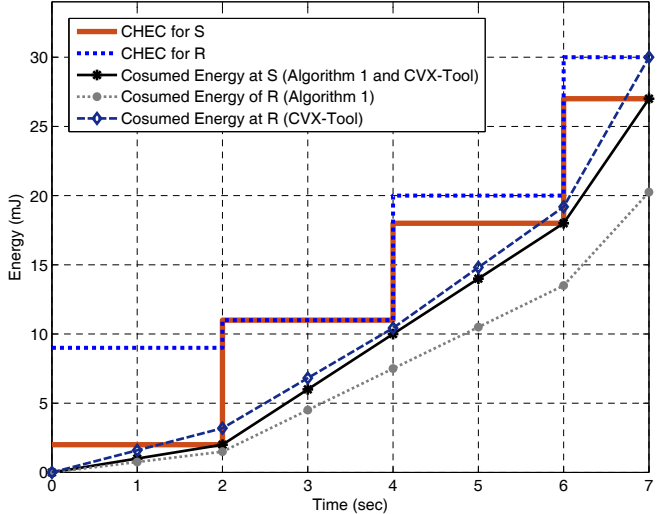}
\caption{Comparison between the allocated powers for S and R in our greedy algorithm and that of the CVX-Tool in Example 1.}
\label{greedy vs cvx}
\end{figure}

\textbf{Example 1:} Suppose that in time instants 0, 2, 4 and 6 sec, the S and R harvest energies $\mathbf{E_1}=[2,\,9,\,7,\,9]$ mJ and $\mathbf{E_2}=[9,\,2,\,9,\,10]$ mJ, respectively. We assume $a=b=2,\,T=7$.
Also, we use \emph{CVX-Tool}, a package for solving disciplined convex programs \cite{cvx} to solve this problem numerically. The CVX-Tool allocates the power for S and R as $\mathbf{P_1^*}=[1,\,4,\,9]$ mW with duration $\mathbf{L_1^*}=[2,\,4,\,1]$ and $\mathbf{P_2^*}=[1.5970,\,3.6188,\,4.3775,\,4.3795,\,10.8113]$ mW with duration $\mathbf{L_2^*}=[2,\,2,\,1,\,1,\,1]$.
Algorithm \ref{algorithm special 1}, on the other hand, assigns $\mathbf{P_1^*}=[1,\,4,\,9]$ mW with duration $\mathbf{L_1^*}=[2,\,4,\,1]$ and $\mathbf{P_2^*}=[0.75,\,3,\,6.75]$ mW with duration $\mathbf{L_2^*}=[2,\,4,\,1]$ for S and R, respectively. The above two power allocations provide the same total transmitted bits with the difference that Algorithm \ref{algorithm special 1} uses the least possible power for the relay. This causes to leave some energies unused ($E_{\texttt{excess}}=9.75$ mJ in this example). This excess energy can be stored in the R for future use or sending its own data in cooperative scenarios. 
The harvested energies and the consumed energies of the S and R in our proposed greedy algorithm and those provided in CVX-Tool are shown in Fig.~\ref{greedy vs cvx}.

Example 1 shows the following Lemma. 

\begin{lemma} \label{cor: lemma 3 not applicable}
Under the optimal policy of NC-EH-$\mathcal{RC}$, if powers of S or R changes in an instant, the total harvested energy in the previous epochs of that node has not necessarily been consumed completely by this instant. Thus,{\cite[Lemma 3]{yang}} does not hold for optimal policy of NC-EH-$\mathcal{RC}$. 
\end{lemma}

\begin{corollary}
The power allocation for S and R based on the \emph{disjoint optimization}, which follows the separate shortest-path algorithm for S and R, constructs a sub-optimal solution for NC-EH-$\mathcal{RC}$. 
\end{corollary}

\begin{IEEEproof}
This follows directly using Lemma~\ref{cor: lemma 3 not applicable}. In other word, {\cite[Lemma 3]{yang}} which is a necessary condition for optimality of disjoint optimization algorithm (shortest path) in S and R, does not hold in general.
\end{IEEEproof}

\begin{remark} \label{remark: lemma 1 not applicable}
In section \ref{numerical results}, we give an example where in the optimal allocation policy, the power of nodes are not monotonically increasing. Therefore, {\cite[Lemma 1]{yang}} does not hold for the general optimal solution of NC-EH-$\mathcal{RC}$.
\end{remark}

\begin{remark}
The disjoint optimization for the S and R is optimal for the special cases studied in \cite{IWCIT}. Those cases that are presented for coherent DF Gaussian RC can be used for NC-EH-$\mathcal{RC}$, as well. 
\end{remark}

In the following, we propose a suboptimal algorithmic solutions for the power allocation problem in NC-EH-$\mathcal{RC}$. This solution is optimal for some EH realizations of S and R. We present some numerical examples in section \ref{numerical results} to study the scenarios in which this suboptimal solution is optimal.

\subsection{Suboptimal Algorithmic Solution for NC-EH-$\mathcal{RC}$ with No ET}\label{opt tx policy}
Here, we come up with a novel approach to find a suboptimal power allocation by introducing a new constraint on the total transmission powers of the nodes. 
First, we consider a simplified problem, which its solution provides a suboptimal solution to our problem. Then, we propose an algorithm which solves the simplified problem.

This suboptimal power allocation for NC-EH-$\mathcal{RC}$ is provided in Algorithm~\ref{alg. tot subopt}.
%\begin{IEEEproof} 
Unlike Algorithm~\ref{algorithm special 1}, this algorithm solves the problem by assuming that the cost function is dominated by the first term under the minimum of (\ref{noncoherent}). If this assumption does not hold, the solution will be suboptimal. We add the following constraint to the problem by combining (\ref{Scausal}) and (\ref{Rcausal}): 
\begin{equation}
\sum\limits_{i = 1}^k {\tilde P_t^i{l^i} \le \sum\limits_{i = 0}^{k - 1} {\tilde E_t^i} } ,\qquad \forall k, \label{Tcausal}
\end{equation}
where $\tilde P_t^i=P_1^i+b^2P_2^i,\,\,\forall i$ and $\tilde E_t^i=E_1^i+b^2E_2^i,\,\,\forall i$.

Therefore, the problem is as follows

\begin{IEEEeqnarray}{l}
\mathop {\max }\limits_{{P_1},{P_2}} \sum\limits_{i = 1}^K \mathcal{C}\left( {\frac{{{P_1^i} + {b^2}{P_2^i}}}{N_0}} \right)l^i \label{new equ}\\
s.t.\qquad(\ref{nonneg})-(\ref{Rcausal}),(\ref{Tcausal}) \nonumber
\end{IEEEeqnarray}

By relaxing constraints (\ref{Scausal}) and (\ref{Rcausal}) from the above problem, we get

\begin{IEEEeqnarray*}{l}
\mathop {\max }\limits_{{\tilde P_t}} \sum\limits_{i = 1}^K {C\left(\frac{\tilde P_t^i}{N_0}\right)}l^i \\ 
s.t.\;\;\; \tilde P_t^i \ge 0, (\ref{Tcausal})
\end{IEEEeqnarray*}

\begin{figure}[tb]
\centering
\includegraphics[width=1\linewidth]{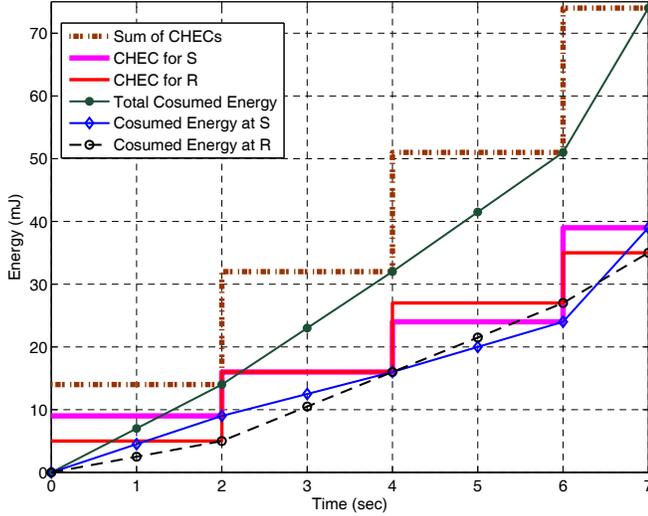}
\caption{An example with modified shortest path power allocation for S and R. Total consumed energy follows the shortest path.}
\label{Total_Subopt_Plot}
\end{figure}

It can be easily seen that the solution for $\mathbf{\tilde P_t}$ in the above problem follows the shortest-path algorithm. 
Now, to find the solution to the problem (\ref{new equ}), it suffices to allocate $P_1^{i*},\,P_2^{i*},\,\forall i$ satisfying (\ref{nonneg})-(\ref{Rcausal}) and $\tilde P_t^{i*}=P_1^{i*}+b^2P_2^{i*},\,\,\forall i$.
One solution for this problem is presented in Algorithm~\ref{alg. tot subopt}. 
Note that if at any time instant, (\ref{Tcausal}) is satisfied with equality, then (\ref{Scausal}) and (\ref{Rcausal}) should also be satisfied with equality. In other words, if the total harvested energies of the network is completely used up in an instant, the same should be happened for energies of S and R. Hence, we  force the S and R to empty their batteries whenever the (\ref{Tcausal}) is active. Within such instants, we allocate power for S and R individually based on shortest path algorithm. This allocation for S and R is called \emph{modified shortest path}. It is obvious that the power allocations for S and R are surely feasible. Fig.~\ref{Total_Subopt_Plot} shows a typical example, in which $\mathbf{P_1}$ and $\mathbf{P_2}$ are obtained using Algorithm~\ref{alg. tot subopt}. In section \ref{numerical results}, we present some examples that this suboptimal solution is optimal.
%This completes the proof.
%\end{IEEEproof}

\begin{algorithm}
\caption{Suboptimal power allocation for NC-EH-$\mathcal{RC}$ with no ET}
\label{alg. tot subopt}
\begin{algorithmic}
\STATE \emph{Total Power Allocation}
\STATE (1) Merge the harvested energies of S and R to produce total harvested energy as $\tilde E_t^i=E_1^i+b^2E_2^i,\,\,i=1,...,K+1$. 
\STATE (2) Find optimal total power allocation as

${o_v} = \mathop {\argmin }\limits_{{o_{v - 1}} < i \le K+1} \frac{{\sum\nolimits_{j = {o_{v - 1}}}^{i-1} {\tilde E_t^j} }}{{{t^i} - {t^{{o_{v - 1}}}}}}$

${ {\tilde {P_t^v}}}^* = \frac{{\sum\nolimits_{j = {o_{v - 1}}}^{{o_v} - 1} {\tilde E_t^j} }}{{{t^{{o_v}}} - {t^{{o_{v - 1}}}}}}$

\STATE (3) Partition transmission time into time slots that total power is fixed (or (\ref{Tcausal}) is active), i.e., $s_j, \,\, j=1,...,Q$ ($\sum\limits_{j=1}^{Q}s_j=T$).
\FOR{$j=1$ \TO $Q$}
\STATE \emph{Individual Power Allocation for S and R}
\STATE (4) Allocate power for S in time slot $s_j$ according to single-user shortest path algorithm.
\STATE (5) Allocate power for R in time slot $s_j$ according to single-user shortest path algorithm.
\ENDFOR
\end{algorithmic}
\end{algorithm}  

%\textbf{Example 2:} Consider a coherent DF Gaussian RC with $a=10$, $b=2$, and $N=1$. For EH pattern, assume at time instants 0, 2, 4 and 6 sec, S and R harvest $E_1=[9,\,\,7,\,\,8,\,\,15]$ mJ and $E_2=[5,\,\,11,\,\,11,\,\,8]$ mJ, respectively, and we have $T=7$sec (see Fig.~\ref{total optimal coherent}). In this case, the total power based near-optimal solution presented in Algorithm \ref{alg. tot subopt} allocates the power of S as $P_1=[4.5,\,\,3.5,\,\,4,\,\,15]$ mW with durations $l_1=[2,\,\,2,\,\,2,\,\,1]$ sec. It allocates power for R as $P_2=[2.5,\,\,5.5,\,\,8]$ mW with durations $l_2=[2,\,\,4,\,\,1]$ sec. This allocation is exactly equal to the power allocated by optimal solution computed numerically using Cvx-Tool. As we see in Fig.~\ref{total optimal coherent}, in optimal solution, monotonic increasing property does not hold as indicated in Corollary \ref{cor: lemmas 2,3 not applicable}. In this example, disjoint-optimal allocation algorithm that allocates the monotonically increasing power sequence to S, i.e., $P_1=[4,\,\,15]$ mW with durations $l_1=[6,\,\,1]$ sec, is not optimal.
%
%\begin{figure}[tb]
%\centering
%\includegraphics[width=1\linewidth]{./total_optimal_coherent.eps}
%\caption{An example that the total power based near-optimal solution is optimal for noncoherent DF Gaussian RC. We assume $a=2$,$b=2$,$N=1$ in this example. }
%\label{total optimal coherent}
%\end{figure}

\section{NC-EH-$\mathcal{RC}$ with One-way Energy Transfer from S to R} \label{sec: one-way}
\begin{figure}[tb]
\centering
\includegraphics[width=1\linewidth]{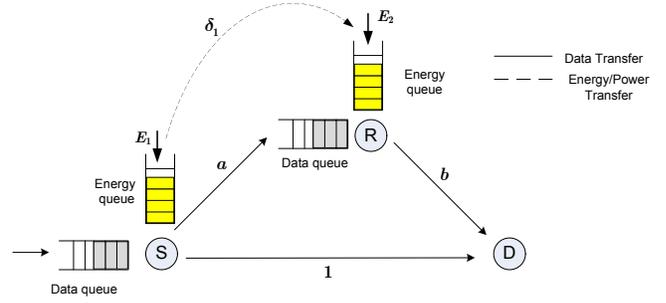}
\caption{Full-duplex RC with energy harvesting S and R and one-way ET from S to R.}
\label{energy trnasfer}
\end{figure}

In this section, we concentrate on NC-EH-$\mathcal{RC}$ with one-way ET from S to R, as in \cite{Gurakan}, and studying the optimal algorithmic solutions for power allocation problem. The system model in this case is depicted in Fig.~\ref{energy trnasfer}. 
We first formulate the problem as follows:
\begin{equation}
\!\!\!\!\!\!\!\!\!\!\mathop {\max }\limits_{\{ P_1^i\} \{ \tilde P_2^i\}\{ \delta_1^i\} } \sum\limits_{i = 1}^{K + 1} {\left( {{\lambda ^i}\mathcal{C}\left(\frac{P_1^i+\tilde P_2^i}{N_0}\right) + (1 - {\lambda ^i})\mathcal{C}\left(\frac{{{[a^2]^\dag}}P_1^i}{N_0}\right)} \right){l^i}} \label{main 1way}
\end{equation}
\begin{IEEEeqnarray}{rll}
\!\!\!s.t.\;\;\;\;&\!\!\! P_1^i \ge 0,\, \tilde P_2^i \ge 0,&\; i = 1,...,K + 1,\\
&{\delta_1^i} \ge 0&\; i = 1,...,K + 1,\label{positive delta}\\ 
&\!\!\!\sum\limits_{i = 1}^k {P_1^i{l^i} \le \sum\limits_{i = 0}^{k - 1} {{E_1^i}-{\delta_1^i}} },&\; k = 1,...,K + 1,\label{eq: causal-S good S}\\
&\!\!\!\sum\limits_{i = 1}^k {\tilde P_2^i{l^i} \le \sum\limits_{i = 0}^{k - 1} {{\tilde E_2^i}+{\delta_1^i}} },&\; k = 1,...,K + 1,\label{eq: causal-R good S}
\end{IEEEeqnarray}
where ${\delta_1^i}$ denotes the energy transfer at epoch $i$ from S to R, $\mathbf{\tilde{P_2}}=b^2\mathbf{P_2}$ and $\mathbf{\tilde{E_2}}=b^2\mathbf{E_2}$. Then, we modify the cost function as 
\begin{equation}
C \ge \min \left\{ {\mathcal{C}\left( {\frac{\tilde P_t}{N_0}} \right),\mathcal{C}\left( {\frac{{{[a^2]^\dag} {P_1}}}{N_0}} \right)} \right\}
\end{equation}
where $\mathbf{\tilde P_t}=\mathbf{P_1}+\mathbf{\tilde{P_2}}$.

\begin{lemma}\label{lemm rep 1way}
We can replace energy causality constraints for R in (\ref{eq: causal-R good S}), with the following

\begin{equation}
\sum\limits_{i = 1}^k {\tilde P_t^i{l^i} \le \sum\limits_{i = 0}^{k - 1} {\tilde E_t^i} },\,\forall k \label{eq: cas Total}
\end{equation}

\end{lemma}

\begin{IEEEproof}
See Appendix \ref{proof lemm rep 1way}.
\end{IEEEproof}

\begin{remark}
We consider $\mathbf{\tilde P_t}=\mathbf{P_1}+b^2\mathbf{P_2}$ instead of $\mathbf{P_t}=\mathbf{P_1}+\mathbf{P_2}$ to define total power. Note that a feasible $P_1^i$ and $P_2^i$ for $\sum\limits_{i = 1}^k {\tilde P_t^i{l^i} \le \sum\limits_{i = 0}^{k - 1} {\tilde E_t^i} },\,\forall k$ does not necessarily satisfy $\sum\limits_{i = 1}^k { P_t^i{l^i} \le \sum\limits_{i = 0}^{k - 1} { E_t^i} },\,\forall k$. In other words, any feasible partitioning of $\tilde P_t^i$ into $P_1^i$ and $P_2^i$ is not necessarily a feasible partitioning for $P_t^i$.
\end{remark}

Applying Lemma~\ref{lemm rep 1way}, (\ref{main 1way})-(\ref{eq: causal-R good S}) are expressed as 
\begin{equation}
\!\!\!\!\!\!\!\!\!\mathop {\max }\limits_{\{ P_1^i\} \{ \tilde P_t^i\}\{ \delta_1^i\} } \sum\limits_{i = 1}^{K + 1} {\left( {{\lambda ^i}\mathcal{C}\left(\frac{\tilde P_t^i}{N_0}\right) + (1 - {\lambda ^i})\mathcal{C}\left(\frac{{{[a^2]^\dag}}P_1^i}{N_0}\right)} \right){l^i}} \label{eq: noncoh s good eh} 
\end{equation}
\begin{IEEEeqnarray}{rll}
\!\!\!s.t.\;\;\;\;&\!\!\! P_1^i \ge 0,\,\tilde P_t^i-P_1^i \ge 0,\,{\delta_1^i} \ge 0&\; i = 1,...,K + 1,\label{nonneg 1way}\\
&\!\!\!\sum\limits_{i = 1}^k {P_1^i{l^i} \le \sum\limits_{i = 0}^{k - 1} {{E_1^i}-{\delta_1^i}} },&\; k = 1,...,K + 1,\label{eq s causality good eh for s}\\
&\!\!\!\sum\limits_{i = 1}^k {\tilde P_t^i{l^i} \le \sum\limits_{i = 0}^{k - 1} {\tilde E_t^i} },&\; k = 1,...,K + 1,\label{cas tot 1et}
\end{IEEEeqnarray}

It can be easily observed that this problem is convex. Thus, the Lagrangian is computed as

\begin{equation}
\begin{array}{l}
L = \sum\limits_{i = 1}^{K + 1} {\left( {{\lambda ^i}\mathcal{C}\left(\frac{\tilde P_t^i}{N_0}\right) + (1 - {\lambda ^i})\mathcal{C}\left(\frac{{{[a^2]^\dag}}P_1^i}{N_0}\right)} \right){l^i}} \\
\,\,\,\, - \sum\limits_{k = 1}^K {{\xi _k}\left( {\sum\limits_{i = 1}^k {P_1^i{l^i}}  - \sum\limits_{i = 0}^{k - 1} ({E_1^i}-{\delta_1^i}) } \right)} \\
\,\,\,\,\, - \sum\limits_{k = 1}^K {{\mu _k}\left( {\sum\limits_{i = 1}^k {\tilde P_t^i{l^i}}  - \sum\limits_{i = 0}^{k - 1} {\tilde E_t^i} } \right)} \\
\,\,\,\, + \sum\limits_{i = 1}^{K + 1} {{\vartheta _i}P_1^i  }+ \sum\limits_{i = 1}^{K + 1} {{\eta _i}\left( \tilde P_t^i-P_1^i\right)}+ \sum\limits_{i = 1}^{K + 1} {{\phi _i}\delta_1^i  }. 
\end{array} \label{equation: lagrangian one-way}
\end{equation}

KKT optimality conditions for (\ref{equation: lagrangian one-way}) are

\begin{IEEEeqnarray}{rrl}
\frac{{(1 - {\lambda ^i}){{[a^2]^\dag}}}}{{{{[a^2]^\dag}}P_1^i+N_0}} - \sum\limits_{k = i}^K {{\xi _k}}  + {\vartheta ^i} - {\eta ^i} &\, = \,& 0,\;\;\;\forall i \label{eq kkt good s}\\
\frac{{{\lambda ^i}}}{{\tilde P_t^i+N_0}} - \sum\limits_{k = i}^K {{\mu _k}}  + {\eta _i} &\, =\, & 0,\;\;\;\forall i\\
 - \sum\limits_{k = i}^K {{\xi _k}}  + {\phi ^i} &\, =\, & 0,\;\;\;\forall i \label{eq kkt 3 s good}
\end{IEEEeqnarray}

with the following complementary slackness conditions

\begin{IEEEeqnarray}{rrl}
{{\mu _k}\left( {\sum\limits_{i = 1}^k {\tilde P_t^i{l^i}}  - \sum\limits_{i = 0}^{k - 1} {\tilde E_t^i} } \right)} &\, = \,& 0,\;\;\;\forall k\\
{{\xi _k}\left( {\sum\limits_{i = 1}^k {P_1^i{l^i}}  - \sum\limits_{i = 0}^{k - 1} ({E_1^i}-{\delta_1^i}) } \right)} &\, =\, & 0,\;\;\;\forall k \label{eq slack good s}\\
{{\vartheta _i}P_1^i  } &\, =\, & 0,\;\;\;\forall k\\
{{\eta _i}\left( \tilde P_t^i-P_1^i\right)} &\, =\, & 0,\;\;\;\forall k\\
{{\phi _i}\delta_1^i  } &\, =\, & 0,\;\;\;\forall k
\end{IEEEeqnarray}

The results for optimal allocated powers are as follows

\begin{equation}
\tilde P_t^{i*} = \frac{{{\lambda ^i}}}{{\sum\limits_{k = i}^K {{\mu _k}} }} - N_0,\,\,\,\forall i \label{opt tot in S good EH}
\end{equation}
and
\begin{equation}
P_1^{i*} = \frac{{(1 - {\lambda ^i})}}{{\sum\limits_{k = i}^K {{\xi _k}} }} - \frac{N_0}{{{{[a^2]^\dag}}}},\,\,\,\,\,\forall i
\end{equation}
and finally
\begin{equation}
P_2^{i*} = \frac{{\tilde P_t^{i*} - P_1^{i*}}}{{{b^2}}},\,\,\,\forall i.
\end{equation}

Considering (\ref{nonneg 1way}), the allocated powers for nodes must be nonnegative. However, there is no incentive to let $P_1^i=0$ or $P_2^i=0$ for any $i$. This is due to the fact that  $E_1^1>0$ and $E_2^1>0$ (see \cite{ozel_jsac} for more details). Thus, the complementary slackness conditions (${\vartheta _i}P_1^i=0,\,\forall i$ and ${{\eta _i}\left( \tilde P_t^i-P_1^i\right)}=0,\,\forall i$) dictates $\vartheta _i=\eta _i=0,\,\forall i$.

Now, we assume that S is in good EH condition. 

\begin{defi}\label{def 2}

We define the class of problem (\ref{main 1way})-(\ref{eq: causal-R good S}) with \emph{S in good EH condition} as the problems, where $P_1^{i*},\,P_2^{i*},\,\forall i$ (in optimal solution) satisfy
\begin{align}
\tilde C_2^i({P_1^i}^*) \ge \tilde C_1^i({P_1^i}^*,{P_2^i}^*),\,\forall i \label{def2 1}
\end{align}
\begin{align}
\forall k,\,\,\,\,\exists \,\,{\delta_1 ^i}^* \ge 0\,\,\mathrm{s.t.}\,\,\,\,\,\sum\limits_{i = 1}^k {{P_1^i}^*{l^i}}  \le \sum\limits_{i = 0}^{k - 1} {E_1^i}  - {\delta_1 ^i}^* \label{def2 2}
\end{align}
This means that S not only scavenged sufficient energy to use for its transmission, but also it can provide energy for R by transferring some parts of its harvested energy.  
\end{defi}

Here, MAC bound ($\tilde C_1$) is the bottleneck. 
The optimal allocation, for the case when S is in good EH condition, is given in Algorithm \ref{algorithm special 2}. 

%S in good EH condition means that the S harvested sufficient energy that enables it to transfer some parts of its energy toward R so that in optimal point $\tilde C_2^i({P_1^i}^*) \ge \tilde C_1^i({P_1^i}^*,{P_2^i}^*),\,\forall i$ and $\forall k,\,\,\,\,\exists \,\,{\delta ^i}^* \ge 0\,\,\mathrm{s.t.}\,\,\,\,\,\sum\limits_{i = 1}^k {{P_1^i}^*{l^i}}  \le \sum\limits_{i = 0}^{k - 1} {E_1^i}  - {\delta ^i}^*$. Algorithm \ref{algorithm special 2} allocates S's power as ${P_1^i}^*=\frac{{\tilde {P_t^i}}^*}{[a^2]^\dag},\,\forall i$, where ${\tilde {P_t^i}}^*$ is the optimal point-to-point solution for total power of S and R. So, the condition for S to be in good EH condition is as presented in algorithm \ref{algorithm special 2}.

\begin{lemma} \label{lemma alg s2}
Algorithm \ref{algorithm special 2} present the optimal algorithmic solution for power allocation problem of NC-EH-$\mathcal{RC}$ with one-way ET from S to R, when S is in good EH condition.
\end{lemma}

\begin{IEEEproof}
When S is in good EH condition, according to Definition~\ref{def 2}, the cost function is $\tilde C_1(\mathbf{\tilde P_t})$.
Therefore, problem (\ref{eq: noncoh s good eh}) reduces to 

\begin{equation}
\!\!\!\!\mathop {\max }\limits_{\{ \tilde P_t^i\} } \sum\limits_{i = 1}^{K + 1} {\mathcal{C}\left(\frac{\tilde P_t^i}{N_0}\right){l^i}} \label{modefy 1}
\end{equation}
\begin{IEEEeqnarray}{rll}
s.t.\qquad& \tilde P_t^i \ge 0,\,&\; i = 1,...,K + 1,\\
&\sum\limits_{i = 1}^k {\tilde P_t^i{l^i} \le \sum\limits_{i = 0}^{k - 1} {\tilde E_t^i} },&\; k = 1,...,K + 1,\label{modefy 2}
\end{IEEEeqnarray}

This is equivalent to inserting ${\lambda ^i}=1,\,\forall i$ in (\ref{eq: noncoh s good eh}). In this case, substituting ${\lambda ^i}=1,\,\forall i$ in (\ref{opt tot in S good EH}), we have

\begin{equation}
\tilde P_t^{i*} = \frac{1}{{\sum\limits_{k = i}^K {{\mu _k}} }} - N_0,\,\,\,\forall i. \label{deducing lemmas}
\end{equation}
It is clear that the optimal solution for $\mathbf{\tilde P_t}$ follows the shortest-path algorithm (we call it $\mathbf{\tilde P_t^*}$). This is due to the fact that three necessary lemmas for this conclusion \cite[Lemmas 1, 2 and 3]{yang} can be deduced using (\ref{deducing lemmas}). The optimal point is $\mathcal{B}^*= \sum\limits_{i = 1}^{K + 1} {\mathcal{C}({{\tilde {P_t^i}^*} \mathord{\left/ {\vphantom {a b}} \right. \kern-\nulldelimiterspace} {N_0}}){l^i}} $ for $\mathbf{\tilde P_t^*}=\mathbf{P_1^*}+b^2\mathbf{P_2^*}$ and feasible $\mathbf{P_1^*}$ and $\mathbf{P_2^*}$, irrespective of exact values for ${P_1^i}^*$ and ${P_2^i}^*$. 
%We note that if in optimal solution we have $\tilde P_t^i = {[a^2]^\dag}P_1^i,\,\forall i$, then for any ${\lambda ^i},\,\forall i$, the cost function in (\ref{eq: noncoh s good eh}) is ${\mathcal{C}({{\tilde P_t^i} \mathord{\left/ {\vphantom {a b}} \right. \kern-\nulldelimiterspace} {N_0}}){l^i}}$. Therefore, the problem reduces to investigate the feasibility of ${\delta^i} \ge 0,\,\forall i$ such that $\sum\limits_{i = 1}^k {\frac{{\tilde P{{_t^{i*}}}}}{{{[a^2]^\dag}}}{l^i}}  \le \sum\limits_{i = 0}^{k - 1} {E_1^i}  - {\delta ^i},\,\forall k$.  
%In other words, if there is a partitioning for $\tilde P_t$, in which $P_1$ is feasible and $\tilde C_2$ is not less than $\tilde C_1$, the solution will be optimal. If EH process at S prevent it's power from making $\tilde C_2^i \ge \tilde C_1^i,\,\forall i$ and still being feasible, then this algorithm cannot provide the optimal solution. 
We set $\tilde P_t^{i*} = {[a^2]^\dag}P_1^{i*},\,\forall i$, which satisfies (\ref{def2 1}). Substitution in (\ref{def2 2}) yields: $\forall k,\,\,\,\,\exists \,\,{\delta_1^{i*}} \ge 0,\,\forall i$ such that $\sum\limits_{i = 1}^k {\frac{{\tilde P{{_t^{i*}}}}}{{{[a^2]^\dag}}}{l^i}}  \le \sum\limits_{i = 0}^{k - 1} {E_1^i}  - {\delta_1 ^{i*}},\,\forall k$. Therefore, our optimal allocations are as 
\begin{align*}
{P_1^i}^*=\frac{{\tilde {P_t^i}}^*}{[a^2]^\dag},\;\;{P_2^i}^*=\frac{([a^2]^\dag-1){\tilde {P_t^i}}^*}{[a^2]^\dag b^2},\,\,\forall i
\end{align*}
This allocation is feasible due to Definition~\ref{def 2}.
Note that this partitioning makes $\tilde C_1$ and $\tilde C_2$ equal. We remark that when S is in good EH condition, Algorithm~\ref{algorithm special 2} never enters the \textbf{return} line. This means that the \textbf{if} condition is always met. This completes the proof.
\end{IEEEproof}

\begin{algorithm} 
\caption{Optimal power allocation algorithm for NC-EH-$\mathcal{RC}$ with one-way ET from S to R}
\label{algorithm special 2}
\begin{algorithmic}

\STATE \emph{Total Power Allocation}
\STATE (1) Set $\tilde E_t^i=E_1^i+b^2E_2^i,\,\,i=1,...,K+1$. 
\STATE (2) Find optimal power allocation for $\mathbf{\tilde P_t}$ as 

${o_v} = \mathop {\argmin }\limits_{{o_{v - 1}} < i \le {K+1}} \frac{{\sum\nolimits_{j = {o_{v - 1}}}^{i-1} {\tilde E_t^j} }}{{{t^i} - {t^{{o_{v - 1}}}}}}$

${\tilde {P_t^v}}^* = \frac{{\sum\nolimits_{j = {o_{v - 1}}}^{{o_v} - 1} {\tilde E_t^j} }}{{{t^{{o_v}}} - {t^{{o_{v - 1}}}}}}$

\STATE \emph{Feasibility Problem for Power Allocation at S}
\IF {$\forall k,\,\,\,\,\exists \,\,{\delta_1 ^i}^* \ge 0\,\,\mathrm{s.t.}\,\,\,\,\,\sum\limits_{i = 1}^k {\frac{{\tilde P{{_t^i}^*}}}{{[a^2]^\dag}}{l^i}}  \le \sum\limits_{i = 0}^{k - 1} {E_1^i}  - {\delta_1 ^i}^*$}

\STATE \emph{Individual Power Allocation for S and R}
\STATE (3) Find optimal power allocation for S as ${P_1^v}^*=\frac{{\tilde {P_t^v}}^*}{[a^2]^\dag}$
\STATE (4) Find optimal power allocation for R as ${P_2^v}^*=\frac{([a^2]^\dag-1){\tilde {P_t^v}}^*}{[a^2]^\dag b^2}$

\ELSE
\RETURN The algorithmic optimal solution is not known in general.
\ENDIF 

\end{algorithmic}
\end{algorithm} 

\begin{remark}
Consider a situation where the harvested energy at S (i.e., $\mathbf{E_1}$) is sufficiently large, so that optimal solution of (\ref{eq: noncoh s good eh}) is obtained while (\ref{eq s causality good eh for s}) is inactive for all epochs. Using slackness condition in (\ref{eq slack good s}), we reach ${\xi _k}=0,\,\forall k$. Substituting this in KKT optimality condition in (\ref{eq kkt 3 s good}), we obtain ${\phi _i}=0,\,\forall i$. Combining with (\ref{eq kkt good s}), we conclude that ${\lambda ^i}=1,\,\forall i$. Therefore, in this case, the problem (\ref{eq: noncoh s good eh})-(\ref{cas tot 1et}) is transformed to (\ref{modefy 1})-(\ref{modefy 2}). Thus, sufficiently large $\mathbf{E_1}$ is a special case of S in good EH condition as expected and hence Algorithm~\ref{algorithm special 2} is optimal in this case.
\end{remark}

\begin{lemma}
In the optimal solution provided by Algorithm \ref{algorithm special 2}, we have the followings
\begin{equation}
\sum\limits_{i = 1}^{K+1} {{P{{_1^i}^*}}{l^i}}  = \sum\limits_{i = 0}^{K} {E_1^i}  - {\delta_1 ^i}^* \label{lem 1}
\end{equation}
\begin{equation}
\sum\limits_{i = 1}^{K+1} {{P{{_2^i}^*}}{l^i}}  = \sum\limits_{i = 0}^{K} {E_2^i}  + \frac{{\delta_1 ^i}^*}{b^2} \label{lem 2}
\end{equation}
This means that S must completely use up its total harvested energy either for transferring toward R or utilizing it for data transmission. On the other hand, R has to use up total energies received by S and harvested through environment by the end of transmission time. 
\end{lemma}

\begin{IEEEproof}
As proved in Lemma \ref{lemma alg s2}, when S is in good EH condition, the cost function is only expressed in $\tilde C_1$, which is a monotonically increasing function of $\mathbf{P_2}$. If the constraints in (\ref{lem 1}) and (\ref{lem 2}) are satisfied with strict inequalities in the optimal solution, we can increase ${\delta_1 ^K}^*$ without violating (\ref{eq s causality good eh for s}). So, we can increase $P{{_2^{K+1}}^*}$, as well.  With this increment, $\tilde C_1$ increases. This contradicts the optimality of ${P{{_1^i}^*}}$, ${P{{_2^i}^*}}$ and ${{\delta_1 ^i}^*}$.
\end{IEEEproof}

Here, an example is presented that optimal solution is obtained using algorithm \ref{algorithm special 2}.

\begin{figure}[tb]
\centering
\includegraphics[width=1\linewidth]{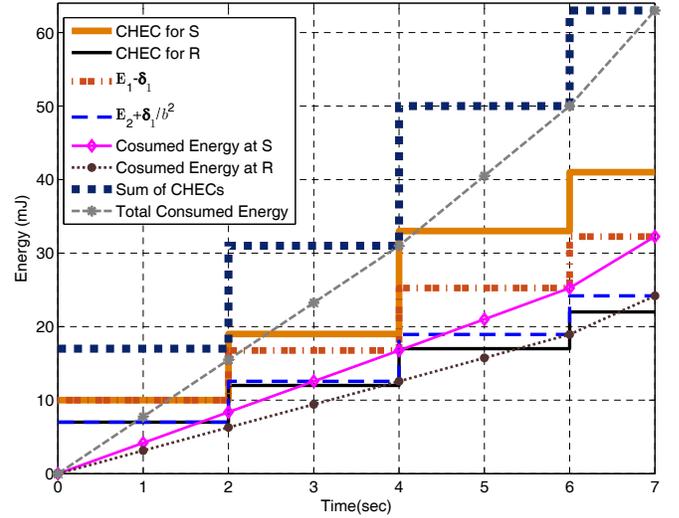}
\caption{Optimal algorithmic solution for the case that one-way ET from S to R is possible.}
\label{example optimal 1way ET}
\end{figure}

\textbf{Example 2:} We assume that S and R harvest $\mathbf{E_1}=[10,\,9,\,14,\,8]$ mJ and $\mathbf{E_2}=[7,\,5,\,5,\,5]$ mJ, respectively at time instants $\mathbf{t}=[0,\,2,\,4,\,6]$ sec. Time duration of interest is $T=7$ sec and we set $a=2$ and $b=2$. This example is the case that there is a positive one-way ET vector $\mathbf{\delta_1^*}=[0,\,2.25,\,5.5,\,1]$ mJ at time instants $\mathbf{t}=[0,\,2,\,4,\,6]$, for which the optimal allocation using algorithm \ref{algorithm special 2} is possible. This algorithm assigns the power of S and R as $\mathbf{P_1}=[4.1875,\,4.25,\,7]$ mW and $\mathbf{P_2}=[3.1406,\,3.1875,\,5.25]$ mW, respectively, with durations $\mathbf{L_1}=\mathbf{L_2}=[4,\,2,\,1]$ sec. Figure \ref{example optimal 1way ET} shows the energy arrivals at S and R and their allocated powers using Algorithm~\ref{algorithm special 2}. Besides, total harvested energy curves and optimum total power, which is allocated based on shortest path algorithm, are shown in this figure. The allocated power of S is restricted to $\mathbf{E_1}-\mathbf{\delta_1^*}$ and that of R is restricted to $\mathbf{E_2}+\mathbf{\delta_1^*}/{b^2}$, as expected.

\section{NC-EH-$\mathcal{RC}$ with Two-way Energy Transfer} \label{sec: two-way}

\begin{figure}[tb]
\centering
\includegraphics[width=1\linewidth]{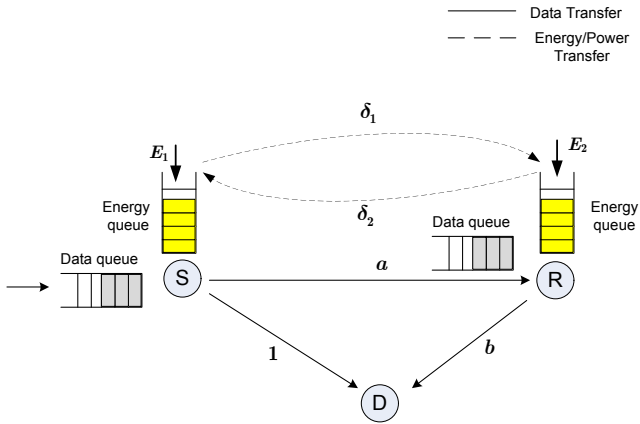}
\caption{Full-duplex RC with two-way ET. S and R share their harvested energies in order to have better control on the network resources.}
\label{biderectional energy cooperation}
\end{figure}

The NC-EH-$\mathcal{RC}$ with two-way ET, in which S and R share their harvested energies with each other is shown in Fig. \ref{biderectional energy cooperation}. This is a case, where we find the general algorithmic optimal solution. 
In this case the problem is as follows
\begin{equation}
\!\!\!\!\!\!\!\!\!\!\mathop {\max }\limits_{{P_1},{\tilde P_2},{\delta_1},\delta_2{}} \sum\limits_{i = 1}^{K+1} {\min } \left\{ { \mathcal{C}\left(\frac{P_1^i+\tilde P_2^i}{N_0}\right),\mathcal{C}\left( {\frac{{{[a^2]^\dag} {P_1^i}}}{N_0}} \right)} \right\}{l^i} \label{equ: 2-ET 1} 
\end{equation}
\begin{IEEEeqnarray}{rll}
\!\!\!\!\!s.t.\qquad&\!\!\! P_1^i \ge 0,\,\tilde P_2^i \ge 0,&\; i = 1,...,K + 1, \label{equ: 2-ET begin}\\
&\!\!\! \delta_1^i \ge 0,\,\delta_2^i \ge 0,&\; i = 1,...,K + 1, \label{equ: 2-ET const 1}\\
&\!\!\! \delta_1^i \cdot \delta_2^i = 0,&\; i = 1,...,K + 1, \label{HD ET}\\
&\!\!\! \sum\limits_{i = 1}^k {P_1^i{l^i} \le \sum\limits_{i = 0}^{k - 1} {E_1^i-\delta_1^i+\delta_2^i} } ,&\; k = 1,...,K + 1, \label{equ: 2-ET end-1}\\
&\!\!\! \sum\limits_{i = 1}^k {\tilde P_2^i{l^i} \le \sum\limits_{i = 0}^{k - 1} {\tilde E_2^i-\delta_2^i+\delta_1^i} } ,&\; k = 1,...,K + 1,\label{equ: 2-ET end}
\end{IEEEeqnarray}
where $\delta_1^i$ and $\delta_2^i$ denote the energy transfers in epoch $i$ in S $\rightarrow$ R and R $\rightarrow$ S directions, respectively.
The constraint (\ref{HD ET}) arises due to the fact that it does not make sense to send and receive energy at the same time. 
We call this constraint \emph{half-duplex energy transfer constraint} in each epoch. 
This problem is not convex due to (\ref{HD ET}). Now, we transform it into a convex optimization problem.

\begin{lemma} \label{lem 2way subs}
The problem in (\ref{equ: 2-ET 1})-(\ref{equ: 2-ET end}) is equivalent to the following convex optimization problem:
\begin{equation}
\!\!\mathop {\max }\limits_{{P_1},{P_2}} \sum\limits_{i = 1}^{K+1} {\min } \left\{ { \mathcal{C}\left(\frac{\tilde P_t^i}{N_0}\right),\mathcal{C}\left( {\frac{{{[a^2]^\dag} {P_1^i}}}{N_0}} \right)} \right\}{l^i} \label{eq subs 1}
\end{equation}
\begin{IEEEeqnarray}{rll}
s.t.\qquad& P_1^i \ge 0,\,\tilde P_t^i \ge 0,&\qquad i = 1,...,K + 1,\\
& \sum\limits_{i = 1}^k {\tilde P_t^i{l^i} \le \sum\limits_{i = 0}^{k - 1} {\tilde E_t^i} } ,&\qquad k = 1,...,K + 1, \label{equ: cas tot 2ET}
\end{IEEEeqnarray}

This means that two problems have the same optimal values. 
\end{lemma}

\begin{IEEEproof}
See Appendix \ref{proof lem 2way subs}.
\end{IEEEproof}

\begin{theorem} \label{theorem 2 ET}
Algorithm \ref{algorithm biderectional} provides the optimal algorithmic solution for the power allocation problem in NC-EH-$\mathcal{RC}$ with two-way ET between S and R, presented in (\ref{equ: 2-ET 1})-(\ref{equ: 2-ET end}).
\end{theorem}

\begin{IEEEproof}[Outline of Proof]
We transform the problem in (\ref{equ: 2-ET 1})-(\ref{equ: 2-ET end}) to the one in (\ref{eq subs 1})-(\ref{equ: cas tot 2ET}), using Lemma~\ref{lem 2way subs}. 
This shows that only feasibility on total power must be met and any ${{P_1^i}^*}$ and ${{P_2^i}^*}$ satisfying (\ref{poof lem 2-et}) are feasible. We set $\tilde P_t^i = {[a^2]^\dag}P_1^i,\,\forall i$. This leads to the following allocations:
\begin{align}
{P_1^v}=\frac{{\tilde {P_t^v}}}{[a^2]^\dag},\;\;{P_2^v}=\frac{([a^2]^\dag-1){\tilde {P_t^v}}}{[a^2]^\dag b^2},\,\,\forall v \label{opt tot gen}
\end{align}
It can be easily seen that in this case the problem is only expressed in terms of $\mathbf{\tilde P_t}$.
Therefore, we first optimally allocate network's total power based on optimal allocation for point-to-point channel (shortest path algorithm). Then, we partition the total power as in (\ref{opt tot gen}). As indicated in the proof of Lemma~\ref{lem 2way subs}, optimal allocation is not unique congruent to the previous parts.
\end{IEEEproof}

\begin{algorithm}
\caption{Optimal power allocation for NC-EH-$\mathcal{RC}$ with two-way ET}
\label{algorithm biderectional}
\begin{algorithmic}
\STATE \emph{Total Power Allocation}
\STATE (1) Set $\tilde E_t^i=E_1^i+b^2E_2^i,\,\,i=1,...,K+1$. 
\STATE (2) Find optimal power allocation for $\mathbf{\tilde P_t}$ as 

${o_v} = \mathop {\argmin }\limits_{{o_{v - 1}} < i \le K+1} \frac{{\sum\nolimits_{j = {o_{v - 1}}}^{i-1} {\tilde E_t^j} }}{{{t^i} - {t^{{o_{v - 1}}}}}}$

${\tilde {P_t^v}}^* = \frac{{\sum\nolimits_{j = {o_{v - 1}}}^{{o_v} - 1} {\tilde E_t^j} }}{{{t^{{o_v}}} - {t^{{o_{v - 1}}}}}}$

\STATE \emph{Individual Power Allocation for S and R}
\STATE (3) Find optimal power allocation for S as ${P_1^v}^*=\frac{{\tilde {P_t^v}}^*}{[a^2]^\dag}$
\STATE (4) Find optimal power allocation for R as ${P_2^v}^*=\frac{([a^2]^\dag-1){\tilde {P_t^v}}^*}{[a^2]^\dag b^2}$

\end{algorithmic}
\end{algorithm}

\begin{figure}[tb]
\centering
\includegraphics[width=1\linewidth]{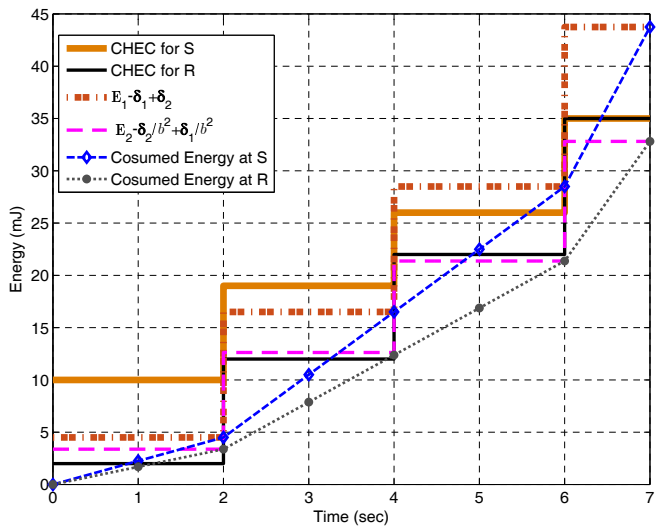}
\caption{Optimal algorithmic solution for NC-EH-$\mathcal{RC}$ with two-way ET in Example 3.}
\label{example optimal 2way ET}
\end{figure}

\textbf{Example 3:} In order to show the performance of Algorithm~\ref{algorithm biderectional}, we assume that S and R harvest energy at time instants $\mathbf{t}=[0,\,2,\,4,\,6]$, with the amounts of $\mathbf{E_1}=[10,\,9,\,7,\,9]$ mJ and $\mathbf{E_2}=[2,\,10,\,10,\,13]$ mJ, respectively. Other parameters are $T=7$ sec, $a=2$ and $b=2$. Algorithm \ref{algorithm biderectional} allocates $\mathbf{P_1}=[2.25,\,6,\,15.25]$ mW and $\mathbf{P_2}=[1.6875,\,4.5,\,11.4375]$ mW with durations $\mathbf{L_1}=\mathbf{L_2}=[2,\,4,\,1]$ sec for S and R, respectively. This needs the energy transfer of $\mathbf{\delta_1^*}=[5.5,\,0,\,0,\,0]$ mJ from S to R and energy transfer of $\mathbf{\delta_2^*}=[0,\,3,\,5,\,6.25]$ mJ from R to S at time instants $\mathbf{t}=[0,\,2,\,4,\,6]$. These values are shown in Fig.~\ref{example optimal 2way ET}. The curves associated with $\mathbf{E_1}-\mathbf{\delta_1^*}+\mathbf{\delta_2^*}$ and $\mathbf{E_2}-\mathbf{\delta_2^*}/{b^2}+\mathbf{\delta_1^*}/{b^2}$ are also shown, which are the upper limits of energy consumption in nodes.

\begin{lemma} \label{end lemma}
General optimal algorithmic solution for nodes in NC-EH-$\mathcal{RC}$ with two-way ET presented in Algorithm~\ref{algorithm biderectional} is equivalent to disjoint optimization for S and R with modified EH pattern $\mathbf{\mathfrak{E}_1}$ and $\mathbf{\mathfrak{E}_2}$, where $\mathbf{\mathfrak{E}_1}=\mathbf{E_1}-\mathbf{\delta_1^*}+\mathbf{\delta_2^*}$ and $\mathbf{\mathfrak{E}_2}=\mathbf{E_2}-\mathbf{\delta_2^*}/{b^2}+\mathbf{\delta_1^*}/{b^2}$.  
\end{lemma}

\begin{IEEEproof}
See Appendix \ref{proof end lemma}.
\end{IEEEproof}

\section{Numerical Results and Discussions} \label{numerical results}
In this section, we investigate the performance of our proposed power allocation algorithms for NC-EH-$\mathcal{RC}$. They consist of optimal and suboptimal solutions that are optimal for some special cases and are presented for NC-EH-$\mathcal{RC}$ with no ET, one-way ET from S to R and two-way ET. 
We consider a band-limited AWGN channel with noise power spectral density ${N_0} = {10^{ - 19}}\,\,{\rm{W/Hz}}$ and bandwidth $W = 1\,\,{\rm{MHz}}$. The distances among nodes are assumed to be 1 Km and path loss is $\psi=100$ dB (typical values used in some EH literatures, e.g. \cite{yang}). Therefore, the Channel to Noise Ratio (CNR) of links in NC-EH-$\mathcal{RC}$ are $\gamma_{sr}=\frac{a^2\psi}{N_0W}$, $\gamma_{rd}=\frac{b^2\psi}{N_0W}$ and $\gamma_{sd}=\frac{\psi}{N_0W}$ for S-R, R-D and S-D links, respectively. The channel gains are set to $a=2$ and $b=2$ for S-R and R-D links, respectively. Harvesting time instants are $\mathbf{t}=[0,\,2,\,4,\,6]$ sec with $T=7$ sec as the time duration of interest.
%Thus, substituting these parameters into the Shannon capacity formula, results in $R = W{\log _2}(1 + p\psi/{N_0}W) = {\log _2}(1 + p/{10^{ - 3}})\,\,{\rm{Mbps}}$ for transmission in the AWGN channel. 
Harvested energies at S and R are samples of Poisson distribution with mean $\bar E_1 =\bar E_2= 10\,{\rm{mJ}}$. 
%Table \ref{tab:coherent} provides total transmitted bits in megabits (Mbits) for the coherent DF Gaussian RC with $a=10$, $b=2$, and $N=1$. We assume that ambient energies are harvested at time instants $t=[0,\,2,\,4,\,6]$ sec and our time duration of interest is $T=7$ sec. EH amounts for S and R are presented in vectors $E_1$ and $E_2$ of Tables \ref{tab:coherent}-\ref{tab:noncoherent-allocation}, respectively. The performance of near-optimal disjoint optimization and total power based algorithm are presented and compared along with optimal solution provided by MATLAB $\mathtt{fmincon}$ function \cite{matlab} for different scenarios: $(c1)$ Scenario 1 is the case that total power based algorithm achieves the same performance of the optimal solution and the better performance compared to that of disjoint optimization. $(c2)$ Scenario 2 is the case that disjoint optimization has the better performance compared to that of the total power based one, in contrast to scenario 1, but transmits 10.1 Kbits less information compared to the optimal solution. $(c3)$ Scenario 3 is the case that two near-optimal solutions achieve the same performance, equal to the optimal value. $(c4)$ Scenario 4 is the case that disjoint optimization algorithm provides the optimal solution and outperforms the total power based allocation algorithm.  

In Table \ref{tab:noncoherent}, the performance of our two proposed suboptimal allocation algorithms (Algorithm~\ref{alg. tot subopt} and disjoint optimization algorithm), and optimal allocation (Algorithm~\ref{algorithm biderectional}) is evaluated and compared with optimal numerical solutions presented by CVX-Tool. Six scenarios are studied: $(s1)$ Scenario 1 is the case that Algorithm~\ref{alg. tot subopt} outperforms the disjoint optimization algorithm, while both of them are suboptimum. $(s2)$ Scenario 2 is the same as $(s1)$ with the difference that Algorithm~\ref{alg. tot subopt} is optimal. $(s3)$ Scenario 3, is the case where disjoint optimization outperform Algorithm~\ref{alg. tot subopt}, while both solutions are suboptimal. $(s4)$ Scenario 4 presents an example that disjoint optimization is optimal and has better performance compared to that of Algorithm~\ref{alg. tot subopt}. $(s5)$ In scenario 5, two algorithms have the same performance, which are suboptimal. $(s6)$ Scenario 6 is the case that both algorithmic solutions are optimal. 
Note that $(s2)$ and $(s3)$  correspond to the examples 2 and 3 in sections \ref{sec: one-way} and \ref{sec: two-way}, respectively. For these scenarios, optimal allocated powers in Algorithms \ref{algorithm special 2} and \ref{algorithm biderectional} are shown in Fig.~\ref{example optimal 1way ET} and Fig.~\ref{example optimal 2way ET}, respectively.

In Table \ref{tab:noncoherent}, the results of optimal CVX-Tool are provided for the cases that one-way ET from S to R and two-way ET between S and R are possible. In $(s1)$, transferring energy from S to R improves the performance but not vice versa. This follows from the fact that the EH at S is better than the EH at R. Thus, energy is required in R more than S. In $(s2)$, we gain nothing by ET (EH at S and R is well equalized and sharing does not improves the performance), whereas in $(s3)$, ET provides the opportunity for better utilization of network energy resources. Here, we achieve better performance, with more capable nodes sharing energy bi-directionally, compared to the no ET and one-way ET cases. In scenarios 4, 5 and 6, unlike $(s1)$, the performance improves only when R is able to transfer some part of its energy toward S. We note that with ET capability added to transmitting nodes, we are able to provide algorithmic optimal solution for the cases that are hard to achieve without them.
We remind that since Algorithm~\ref{algorithm biderectional} presents the optimal solution for the case with two-way ET, it leads to the same result as CVX-Tool (see the last column of Table \ref{tab:noncoherent}), but with different power allocations which are given in Tables \ref{tab:noncoherent-allocation1}, \ref{tab:noncoherent-allocation2}, and \ref{tab:noncoherent-allocation3}.

Tables \ref{tab:noncoherent-allocation1}, \ref{tab:noncoherent-allocation2}, and \ref{tab:noncoherent-allocation3} provide the powers allocated to S and R in different solution methods for scenarios 2, 4, and 6, respectively. In Table \ref{tab:noncoherent-allocation1}, where Algorithm~\ref{alg. tot subopt} is optimal, R's allocated powers are $[3,\,2.5,\,5]$ mJ with durations $[4,\,2,\,1]$ sec. On the other hand, in disjoint optimization (suboptimum in this example), they are $[2.8333,\,5]$ mJ with durations $[6,\,1]$ sec. Therefore, optimal powers do not have necessarily monotonic behaviour. Note that the CVX-Tool shows the same behaviour as Algorithm~\ref{alg. tot subopt}. Unlike $(s2)$, Table \ref{tab:noncoherent-allocation2} shows the case where equalized power of S in disjoint optimal algorithm, i.e., $P_1=5.4286$ mW with duration $L_1=7$ sec, provides optimal performance. This exceeds the performance of Algorithm~\ref{alg. tot subopt} ($\mathbf{P_1}=[5.5,\,5]$ mW with durations $\mathbf{L_1}=[6,\,1]$ sec for S with same power allocation for R). In Table \ref{tab:noncoherent-allocation3}, where two suboptimal allocation algorithms are optimal, power allocations are exactly the same. 
Also, we observe in these tables that even though Algorithm~\ref{algorithm biderectional} provides the same optimal value as the CVX-Tool with two-way ET, its allocated powers are totally different. We see in three scenarios of Tables \ref{tab:noncoherent-allocation1}, \ref{tab:noncoherent-allocation2}, and \ref{tab:noncoherent-allocation3} that the power allocation of Algorithm~\ref{algorithm biderectional} is more fair compared to diverse power allocation of CVX-Tool. Besides, high power allocation for S in CVX-Tool may cause some technical difficulties, if utilized in practice. This highlights the applicability of Algorithm \ref{algorithm biderectional} in practical transmitter schedulers.

%\begin{table*}[th]
%\caption{Total transmitted bits (in Mbits) in the suboptimal disjoint and total power based power allocation algorithms in the coherent DF Gaussian RC with $a=10$, $b=2$, $N=1$, and harvesting time instants $t=[0\;\; 2\;\; 4\;\; 6]$ with $T=7$sec.} % title name of the table
%\centering          % centering table
%\begin{tabular}{l lccr lccr c c c}    % creating 10 columns
%\toprule \hline \\ [-1.5ex]      % inserting double-line
%Scenarios & \multicolumn{4}{c}{EH values for S, ${E_1}$} & \multicolumn{4}{c}{EH values for R, ${E_2}$}  & Total-power based Algorithm & Disjoint optimization  & MATLAB $\mathtt{fmincon}$ function \cite{matlab}
%\\[1ex]  % [1ex] adds vertical space
%\hline \\ [-1.5ex]      
%     % inserts single-line
%% Entering 1st row
%Scenario 1 & [10 & 9 & 7 & 9]& [ 2 & 10 & 10 & 13]& 36.7079 & 36.6501 & 36.7079  \\[1ex] 
%
%% Entering 2nd row
%Scenario 2 & [7 & 10 & 13 & 11]& [ 9 & 12 & 9 & 1]& 37.725 & 38.0693 & 38.0794 \\[1ex] 
%
%% Entering 3rd row
%Scenario 3 & [12 & 14 & 10 & 12]& [ 8 & 13 & 12 & 15]& 40.8942 & 40.8942 & 40.8942  \\[1ex] 
%
%Scenario 4 & [14 & 7 & 11 & 6]& [ 13 & 11 & 19 & 9]& 41.2186 & 41.2181 & 41.2186  \\[1ex] 
%
%% [1ex]adds vertical space
%%\hline % inserts single-line
%\bottomrule
%\end{tabular}
%\label{tab:coherent}
%\end{table*}

%\begin{landscape}
\begin{table*}[th]
\caption{Total transmitted bits (in Mbits) in the suboptimal algorithms, which are proposed for NC-EH-$\mathcal{RC}$. Optimal power allocation using convex optimization tool is included for no-ET, one-way ET and two-way ET. The results of optimal power allocation in algorithm \ref{algorithm biderectional} is also included.} % title name of the table
\centering          % centering table
\begin{tabular}{l lccr lccr c c c c c}    % creating 10 columns
\toprule \hline\\ [-1ex]    % inserting double-line
Scenarios & \multicolumn{4}{c}{\multirow{2}{2cm}{\centering EH values for S, ${\mathbf{E_1}}$}} & \multicolumn{4}{c}{\multirow{2}{2cm}{\centering EH values for R, ${\mathbf{E_2}}$}} & \multirow{-1}{2cm}{\centering Algorithm \ref{alg. tot subopt}} &  \multirow{-1}{2cm}{\centering Disjoint Optimization Algorithm} & \multirow{1}{2cm}{CVX-Tool \cite{cvx}} & \multirow{-1}{2cm}{\centering CVX-Tool with one-way ET} & \multirow{3}{1.7cm}{\centering CVX-Tool with two-way ET, and Algorithm \ref{algorithm biderectional}}
\\[8.5ex]  % [1ex] adds vertical space
\hline\\  [-1ex]    
     % inserts single-line

Scenario 1 & [10 & 21 & 14 & 9]& [ 7 & 5 & 8 & 11]& 31.8339 & 31.8082 & 32.1965 & 32.4212 & 32.4212 \\[1ex] 

Scenario 2 & [10 & 9 & 14 & 8]& [ 7 & 5 & 5 & 5]& 29.7968 & 29.7821 & 29.7968 & 29.7968 & 29.7968 \\[1ex] 

Scenario 3 & [10 & 9 & 7 & 9]& [ 2 & 10 & 10 & 13]& 28.2032 & 28.4398 & 28.9548 & 29.8207 & 31.1735 \\[1ex] 

Scenario 4 & [17 & 7 & 9 & 5]& [ 13 & 7 & 9 & 10]& 31.5337 & 31.5387 & 31.5387 & 31.5387 & 33.6705 \\[1ex]

Scenario 5 & [7 & 11 & 15 & 15]& [ 12 & 15 & 10 & 8]& 32.3543 & 32.3543 & 32.7000 & 32.7000 & 35.3402  \\[1ex]

Scenario 6 & [7 & 11 & 11 & 9]& [ 10 & 7 & 11 & 12]& 31.1175 & 31.1175 & 31.1175 & 31.1175 & 33.4912  \\[1ex]

% [1ex]adds vertical space
%\hline % inserts single-line
\bottomrule
\end{tabular}
\label{tab:noncoherent}
\end{table*}
%\end{landscape}

\begin{table*}[th]
\caption{Power allocation for S and R in NC-EH-$\mathcal{RC}$. The results are presented for scenario 2 of Table \ref{tab:noncoherent}, in which Algorithm \ref{alg. tot subopt} is optimal. Power allocation in Algorithm \ref{algorithm biderectional} is also included to be compared with that of CVX-Tool with two-way ET.} % title name of the table
\centering          % centering table
\begin{tabular}{l l lllccr lllccr}    % creating 10 columns
\toprule\hline\\ [-1ex]    % inserting double-line
Scenario & Solution Methods &  \multicolumn{6}{c}{\centering Power Allocation for S, ${\mathbf{P_1}}$ and ${\mathbf{L_1}}$} & \multicolumn{6}{c}{\centering Power Allocation for R, ${\mathbf{P_2}}$ and ${\mathbf{L_2}}$}  \\[2ex] 
\hline\\  [-1ex]  

\multirow{10}{*}{\centering Scenario 2} &\multirow{2}{*}{\centering Algorithm \ref{alg. tot subopt}}  & ${\mathbf{P_1}}$&=&[4.75 & 7 & 8 & ] & ${\mathbf{P_2}}$&=&[3 & 2.5 & 5 & ] \\[1ex] 
                                       &                                                         & ${\mathbf{L_1}}$&=&[4 & 2 & 1 & ] & ${\mathbf{L_2}}$&=&[4 & 2 & 1 & ] \\[1ex]

                                       &\multirow{2}{*}{\centering Disjoint Optimization Algorithm}        & ${\mathbf{P_1}}$&=&[4.75 & 7 & 8 & ] & ${\mathbf{P_2}}$&=&[2.8333 & 5 &  & ] \\[1ex]
                                       &                                                         &${\mathbf{L_1}}$&=&[4 & 2 & 1 & ] & ${\mathbf{L_2}}$&=&[6 & 1 &  & ] \\[1ex]

                                       &\multirow{2}{*}{\centering CVX-Tool}                     & ${\mathbf{P_1}}$&=& [4.6569 & 4.8430 & 7 & 8] & ${\mathbf{P_2}}$&=&[3.0233 & 2.9767 & 2.5 & 5] \\[1ex]
                                       &                                                         & ${\mathbf{L_1}}$&=&[2 & 2 & 2 & 1] & ${\mathbf{L_2}}$&=&[2 & 2 & 2 & 1] \\[1ex]

                                       &\multirow{2}{*}{\centering CVX-Tool with one-way ET}     & ${\mathbf{P_1}}$&=&[4.6709 & 4.7416 & 6.2051 & 9.7647] & ${\mathbf{P_2}}$&=&[3.0198& 3.0021 & 2.6987 & 4.5588] \\[1ex]
                                       &                                                         & ${\mathbf{L_1}}$&=&[2 & 2 & 2 & 1] & ${\mathbf{L_2}}$&=&[2 & 2 & 2 & 1] \\[1ex]

                                       &\multirow{2}{*}{\centering Algorithm \ref{algorithm biderectional} }  & ${\mathbf{P_1}}$&=&[4.1875 & 4.25 & 7 & ] & ${\mathbf{P_2}}$&=&[3.1406 & 3.1875 & 5.25 & ] \\[1ex]
                                       &                                                         & ${\mathbf{L_1}}$&=&[4 & 2 & 1 & ] & ${\mathbf{L_2}}$&=&[4 & 2 & 1 & ] \\[1ex]

                                       &\multirow{2}{*}{\centering CVX-Tool with two-way ET}  & ${\mathbf{P_1}}$&=&[14.336 & 14.2945 & 14.4894 & 23.7573] & ${\mathbf{P_2}}$&=&[0.6035 & 0.6139 & 0.6276 & 1.0607] \\[1ex]
                                       &                                                         & ${\mathbf{L_1}}$&=&[2 & 2 & 2 & 1] & ${\mathbf{L_2}}$&=&[2 & 2 & 2 & 1] \\[1ex]

% [1ex]adds vertical space
%\hline % inserts single-line
\bottomrule
\end{tabular}
\label{tab:noncoherent-allocation1}
\end{table*}

\begin{table*}[th]
\caption{Power allocation for S and R in NC-EH-$\mathcal{RC}$. The results are presented for scenario 4 of Table \ref{tab:noncoherent}, in which disjoint optimization algorithm is optimal. Power allocation in Algorithm \ref{algorithm biderectional} is also included to be compared with that of CVX-Tool with two-way ET.} % title name of the table
\centering          % centering table
\begin{tabular}{l l lllccr lllccr}    % creating 10 columns
\toprule\hline\\ [-1ex]    % inserting double-line
Scenario & Solution Methods &  \multicolumn{6}{c}{\centering Power Allocation for S, ${\mathbf{P_1}}$ and ${\mathbf{L_1}}$} & \multicolumn{6}{c}{\centering Power Allocation for R, ${\mathbf{P_2}}$ and ${\mathbf{L_2}}$}  \\[2ex] 
\hline\\  [-1ex]  

\multirow{10}{*}{\centering Scenario 4} &\multirow{2}{*}{\centering Algorithm \ref{alg. tot subopt}}  & ${\mathbf{P_1}}$&=&[5.5 & 5 &  & ] & ${\mathbf{P_2}}$&=&[4.8333 & 10 &  & ] \\[1ex] 
                                       &                                                         & ${\mathbf{L_1}}$&=&[6 & 1 &  & ] & ${\mathbf{L_2}}$&=&[6 & 1 &  & ] \\[1ex] 
                                       &\multirow{2}{*}{\centering Disjoint Optimization Algorithm}        & ${\mathbf{P_1}}$&=&[5.4286 &  &  & ] & ${\mathbf{P_2}}$&=&[4.8333 & 10 &  & ] \\[1ex]
                                       &                                                         &${\mathbf{L_1}}$&=&[7 &  &  & ] & ${\mathbf{L_2}}$&=&[6 & 1 &  & ] \\[1ex]
                                       &\multirow{2}{*}{\centering CVX-Tool}                     & ${\mathbf{P_1}}$&=& [5.4288 & 5.4272 &  & ] & ${\mathbf{P_2}}$&=&[4.6786 & 4.7186 & 4.9065 & 10.3926] \\[1ex]
                                       &                                                         & ${\mathbf{L_1}}$&=&[6 & 1 &  & ] & ${\mathbf{L_2}}$&=&[2 & 2 & 2 & 1] \\[1ex]
                                       &\multirow{2}{*}{\centering CVX-Tool with one-way ET}     & ${\mathbf{P_1}}$&=&[5.4286 & &  & ] & ${\mathbf{P_2}}$&=&[4.7325& 4.7533 & 4.8809 & 10.2667] \\[1ex]
                                       &                                                         & ${\mathbf{L_1}}$&=&[7 &  &  & ] & ${\mathbf{L_2}}$&=&[2 & 2 & 2 & 1] \\[1ex]

                                       &\multirow{2}{*}{\centering Algorithm \ref{algorithm biderectional} }  & ${\mathbf{P_1}}$&=&[6.2083 & 11.25 &  & ] & ${\mathbf{P_2}}$&=&[4.6563 & 8.4375 &  & ] \\[1ex]
                                       &                                                         & ${\mathbf{L_1}}$&=&[6 & 1 &  & ] & ${\mathbf{L_2}}$&=&[6 & 1 &  & ] \\[1ex]

                                       &\multirow{2}{*}{\centering CVX-Tool with two-way ET}  & ${\mathbf{P_1}}$&=&[21.6191 & 21.5233 & 21.4378 & 38.4547] & ${\mathbf{P_2}}$&=&[0.8036 & 0.8275 & 0.8489 & 1.6363] \\[1ex]
                                       &                                                         & ${\mathbf{L_1}}$&=&[2 & 2 & 2 & 1] & ${\mathbf{L_2}}$&=&[2 & 2 & 2 & 1] \\[1ex]

% [1ex]adds vertical space
%\hline % inserts single-line
\bottomrule
\end{tabular}
\label{tab:noncoherent-allocation2}
\end{table*}

\begin{table*}[th]
\caption{Power allocation for S and R in NC-EH-$\mathcal{RC}$. The results are presented for scenario 6 of Table \ref{tab:noncoherent}, in which both of the proposed suboptimal algorithms are optimal. Power allocation in Algorithm \ref{algorithm biderectional} is also included to be compared with that of CVX-Tool with two-way ET.} % title name of the table
\centering          % centering table
\begin{tabular}{l l lllccr lllccr}    % creating 10 columns
\toprule\hline\\ [-1ex]    % inserting double-line
Scenario & Solution Methods &  \multicolumn{6}{c}{\centering Power Allocation for S, ${\mathbf{P_1}}$ and ${\mathbf{L_1}}$} & \multicolumn{6}{c}{\centering Power Allocation for R, ${\mathbf{P_2}}$ and ${\mathbf{L_2}}$}  \\[2ex] 
\hline\\  [-1ex]  

\multirow{10}{*}{\centering Scenario 6} &\multirow{2}{*}{\centering Algorithm \ref{alg. tot subopt}}  & ${\mathbf{P_1}}$&=&[3.5 & 5.5 & 9 & ] & ${\mathbf{P_2}}$&=&[4.25 & 5.5 & 12 &] \\[1ex] 
                                       &                                                         & ${\mathbf{L_1}}$&=&[2 & 4 & 1 & ] & ${\mathbf{L_2}}$&=&[4 & 2 & 1 & ] \\[1ex] 
                                       &\multirow{2}{*}{\centering Disjoint Optimization Algorithm}        & ${\mathbf{P_1}}$&=&[3.5 & 5.5 & 9 & ] & ${\mathbf{P_2}}$&=&[4.25 & 5.5 & 12 &] \\[1ex]
                                       &                                                         &${\mathbf{L_1}}$&=&[2 & 4 & 1 & ] & ${\mathbf{L_2}}$&=&[4 & 2 & 1 & ] \\[1ex]
                                       &\multirow{2}{*}{\centering CVX-Tool}                     & ${\mathbf{P_1}}$&=& [3.5 & 5.4989 & 5.5011 & 9] & ${\mathbf{P_2}}$&=&[3.3394 & 4.8254 & 5.4913 & 12.6878] \\[1ex]
                                       &                                                         & ${\mathbf{L_1}}$&=&[2 & 2 & 2 & 1] & ${\mathbf{L_2}}$&=&[2 & 2 & 2 & 1] \\[1ex]
                                       &\multirow{2}{*}{\centering CVX-Tool with one-way ET}     & ${\mathbf{P_1}}$&=&[3.5 & 5.4998 & 5.5002 & 9] & ${\mathbf{P_2}}$&=&[3.3420& 4.8907 & 5.4962 & 12.5423] \\[1ex]
                                       &                                                         & ${\mathbf{L_1}}$&=&[2 & 2 & 2 & 1] & ${\mathbf{L_2}}$&=&[2 & 2 & 2 & 1] \\[1ex]

                                      &\multirow{2}{*}{\centering Algorithm \ref{algorithm biderectional} }  & ${\mathbf{P_1}}$&=&[5.3750 & 6.8750 & 14.25 & ] & ${\mathbf{P_2}}$&=&[4.0313 & 5.1563 & 10.6875 & ] \\[1ex]
                                       &                                                         & ${\mathbf{L_1}}$&=&[4 & 2 & 1 & ] & ${\mathbf{L_2}}$&=&[4 & 2 & 1 & ] \\[1ex]

                                       &\multirow{2}{*}{\centering CVX-Tool with two-way ET}  & ${\mathbf{P_1}}$&=&[18.5759 & 18.5001 & 23.6885 & 48.8493] & ${\mathbf{P_2}}$&=&[0.7310 & 0.75 & 0.9529 & 2.0377] \\[1ex]
                                       &                                                         & ${\mathbf{L_1}}$&=&[2 & 2 & 2 & 1] & ${\mathbf{L_2}}$&=&[2 & 2 & 2 & 1] \\[1ex]

% [1ex]adds vertical space
%\hline % inserts single-line
\bottomrule
\end{tabular}
\label{tab:noncoherent-allocation3}
\end{table*}

\section{Conclusion}\label{sec:coclusion}
We investigated the optimal power allocation for a three-node full-duplex non-coherent decode-and-forward Gaussian relay channel with energy harvesting source and relay nodes (called NC-EH-$\mathcal{RC}$). Three cases were considered based on the capability of the source and the relay nodes to transfer parts of their energies to each other, namely no ET, one-way ET and two-way ET. The original problem for NC-EH-$\mathcal{RC}$ with no ET has a complicated min-max form, which is not easy to solve. We showed that it is transformed to a tractable convex optimization problem, which can be solved efficiently. However, convex optimization did not provide any structural property of optimal solution to be used in devising algorithmic solutions. Following a different perspective, we studied cases where optimal algorithmic solutions are found. These cases were investigated to give insight by revealing some key specifications of general optimal solution. These specifications discriminated our problem from the others in the existing works and showed that our problem can not be reduced to the existing EH problems. Also, we proposed some suboptimal algorithmic solutions that are optimal for some realizations of EH pattern at S and R. Moreover, in NC-EH-$\mathcal{RC}$ with one-way ET, we found a class of problems, where the optimal algorithmic solution was devised. For NC-EH-$\mathcal{RC}$ with two-way ET, we derived some interesting properties of optimal solution that are used to find optimal algorithmic solution \emph{in general}.
Besides, the performance of our proposed algorithms were evaluated numerically and compared with optimal numerical convex optimization tools. Numerical results highlighted the applicability of our proposed algorithms in practical transmitter schedulers.

\appendices

\section{Proof of Lemma \ref{lemm rep 1way}} \label{proof lemm rep 1way}
It is obvious that for a feasible solution satisfying (\ref{eq: causal-S good S}) and (\ref{eq: causal-R good S}), (\ref{eq: cas Total}) is also held by combining (\ref{eq: causal-S good S}) and (\ref{eq: causal-R good S}). To show the converse, if (\ref{eq: cas Total}) is satisfied for an arbitrary $i=\breve{i}$, then 
\begin{equation*}
\sum\limits_{i = 1}^{\breve{i}} {({P_1^i+\tilde P_2^i}){l^i} \le \sum\limits_{i =1 }^{\breve{i} } {E_1^{i-1}+\tilde E_2^{i-1}} },
\end{equation*}

Subtracting $P_1^i{l^i}>0,\,\forall i$ from both sides of the above inequality, we have

\begin{equation*}
\sum\limits_{i = 1}^{\breve{i}} {{\tilde P_2^i}{l^i} \le \sum\limits_{i = 1}^{\breve{i}} {\tilde E_2^{i-1}+E_1^{i-1}-P_1^i l^i} },
\end{equation*}

Since $E_1^{i-1}-P_1^i l^i \ge 0$ according to (\ref{positive delta})-(\ref{eq: causal-S good S}), defining $\delta^i=E_1^i-P_1^i l^i$, we reach (\ref{eq: causal-R good S}). 

\section{Proof of Lemma \ref{lem 2way subs}} \label{proof lem 2way subs}
It suffices to show that equations (\ref{equ: 2-ET const 1})-(\ref{equ: 2-ET end}) can be replaced with (\ref{equ: cas tot 2ET}). The direct proof is straightforward as we reach to (\ref{equ: cas tot 2ET}) by combining (\ref{equ: 2-ET end-1})-(\ref{equ: 2-ET end}). Then, (\ref{equ: 2-ET const 1})-(\ref{HD ET}) can be omitted as they are irrelevant to cost function and other constraints. To prove the converse, using (\ref{equ: cas tot 2ET}) we have 

\begin{equation}
\sum\limits_{i = 1}^k {(P_1^i+\tilde P_2^i){l^i} \le \sum\limits_{i = 1}^{k } {(E_1^{i-1}+\tilde E_2^{i-1})} } ,\qquad k = 1,...,K + 1, \label{poof lem 2-et}
\end{equation}

Subtracting $\tilde P_2^i{l^i} \ge 0$ from both sides, results in
\begin{equation}
\sum\limits_{i = 1}^k {P_1^i{l^i} \le \sum\limits_{i = 1}^{k } {E_1^{i-1}+{\Delta}^i} } ,\qquad k = 1,...,K + 1, \label{pr lm 1}
\end{equation}
where ${\Delta}^i=\tilde E_2^{i-1}-\tilde P_2^i{l^i},\,\forall i$.
This yields 
\begin{equation}
\sum\limits_{i = 1}^k {\tilde P_2^i{l^i} = \sum\limits_{i = 1}^{k } {\tilde E_2^{i-1}-{\Delta}^i} } ,\qquad k = 1,...,K + 1, \label{pr lm 2}
\end{equation}
which can be expressed as
\begin{equation}
\sum\limits_{i = 1}^k {\tilde P_2^i{l^i} \le \sum\limits_{i = 1}^{k } {\tilde E_2^{i-1}-{\Delta}^i} } ,\qquad k = 1,...,K + 1, \label{pr lm }
\end{equation}
Now, if $\Delta^i>0$, we define $\Delta^i=\delta_1^i$; otherwise, for $\Delta^i<0$, we define $\Delta^i=-\delta_2^i$. Therefore, we have (\ref{equ: 2-ET const 1})-(\ref{equ: 2-ET end}). It is obvious that $\delta_1^i,\,\delta_2^i,\,\forall i$ are not unique. This completes the proof.
%On the other hand, subtracting $P_1^i{l^i} \ge 0$ from both sides of (\ref{poof lem 2-et}), we have
%\begin{equation}
%\sum\limits_{i = 1}^k {\tilde P_2^i{l^i} \le \sum\limits_{i = 1}^{k } {\tilde E_2^{i-1}+{\Delta''}^i} } ,\qquad k = 1,...,K + 1, \label{pr lm }
%\end{equation}
%where ${\Delta''}^i= E_1^{i-1}- P_1^i{l^i},\,\forall i$.
%We can find ${\Delta}^i$ such that (1): ${\Delta''}^i \le{\Delta}^i \le -{\Delta'}^i$ or (2): ${\Delta'}^i \le{\Delta}^i \le -{\Delta''}^i$.!!!!
%For (1), we get
%\begin{equation*}
%\sum\limits_{i = 1}^k {P_1^i{l^i} \le \sum\limits_{i = 1}^{k } {E_1^{i-1}-{\Delta}^i} } ,\qquad k = 1,...,K + 1, 
%\end{equation*}
%\begin{equation*}
%\sum\limits_{i = 1}^k {\tilde P_2^i{l^i} \le \sum\limits_{i = 1}^{k } {\tilde E_2^{i-1}+{\Delta}^i} } ,\qquad k = 1,...,K + 1, 
%\end{equation*}
%and for (2)
%\begin{equation*}
%\sum\limits_{i = 1}^k {P_1^i{l^i} \le \sum\limits_{i = 1}^{k } {E_1^{i-1}+{\Delta}^i} } ,\qquad k = 1,...,K + 1, 
%\end{equation*}
%\begin{equation*}
%\sum\limits_{i = 1}^k {\tilde P_2^i{l^i} \le \sum\limits_{i = 1}^{k } {\tilde E_2^{i-1}-{\Delta}^i} } ,\qquad k = 1,...,K + 1, 
%\end{equation*}
%In order to make (\ref{pr lm 1})-(\ref{pr lm 2}) equivalent to (\ref{equ: cas tot 2ET}), we should have $\Delta_1^i+\Delta_2^i=0,\,\forall i$. This leads to $\Delta_1^i=-\Delta_2^i,\,\forall i$.
\section{Proof of Lemma \ref{end lemma}} \label{proof end lemma}
In optimal solution of Algorithm~\ref{algorithm biderectional}, we have $\tilde C_1=\tilde C_2$. Therefore, the problem is

\begin{IEEEeqnarray}{l}
\mathop {\max }\limits_{{P_1},{\tilde P_2},{\delta_1},\delta_2{}} \sum\limits_{i = 1}^{K+1}{\mathcal{C}\left( {\frac{{{[a^2]^\dag} {P_1^i}}}{N_0}} \right)}{l^i}\\
\qquad s.t.\qquad (\ref{equ: 2-ET begin})-(\ref{equ: 2-ET end}).
\end{IEEEeqnarray}

This is simplified to 

\begin{IEEEeqnarray}{l}
\mathop {\max }\limits_{{P_1},{\delta_1},\delta_2{}} \sum\limits_{i = 1}^{K+1}{\mathcal{C}\left( {\frac{{{[a^2]^\dag} {P_1^i}}}{N_0}} \right)}{l^i}\\
 s.t.\qquad P_1^i \ge 0,\; \forall i, \mathrm{and} \;\;(\ref{equ: 2-ET const 1}),(\ref{HD ET}),(\ref{equ: 2-ET end-1}), 
\end{IEEEeqnarray}
where $P_2^i$ is removed from the problem. Substituting $\mathbf{\delta_1}=\mathbf{\delta_1^*}$ and $\mathbf{\delta_2}=\mathbf{\delta_2^*}$, we have

\begin{IEEEeqnarray}{l}
\mathop {\max }\limits_{{P_1}} \sum\limits_{i = 1}^{K+1}{\mathcal{C}\left( {\frac{{{[a^2]^\dag} {P_1^i}}}{N_0}} \right)}{l^i}\\
 s.t.\qquad P_1^i \ge 0,\; \forall i, \\
\;\; \qquad \sum\limits_{i = 1}^k {P_1^i{l^i} \le \sum\limits_{i = 0}^{k - 1} \mathfrak{E}_1^i } ,\; \forall k, 
\end{IEEEeqnarray}
It is clear that the solution of this problem (convex optimization problem in single variable $\mathbf{P_1}$) is the shortest-path algorithm applied to modified EH pattern $\mathbf{\mathfrak{E}_1}$. Using a similar technique, we can achieve the desired result for $\mathbf{P_2}$. 

%\section*{Acknowledgment}




%\bibliographystyle{IEEEtran}
%\bibliographystyle{wileyj}
\bibliographystyle{IEEEtran}
\bibliography{My_Refs}
%
%\begin{thebibliography}{2}
%\bibitem{terk}
%\end{thebibliography}

\end{document}